\documentclass[fleqn,usenatbib]{mnras}
\usepackage{lmodern}
\usepackage{newtxtext,newtxmath}
\usepackage[T1]{fontenc}

\DeclareRobustCommand{\VAN}[3]{#2}
\let\VANthebibliography\thebibliography
\def\thebibliography{\DeclareRobustCommand{\VAN}[3]{##3}\VANthebibliography}

\usepackage{graphicx}	
\usepackage{amsmath}	
\usepackage{subcaption}
\usepackage{multirow} 
\usepackage{array} 
\usepackage{booktabs}
\usepackage{tabularx}
\usepackage{adjustbox}

\newcommand{\kms}{\,km\,s$^{-1}$}
\newcommand{\nel}{$n_\mathrm{e}\ $}
\newcommand{\te}{$T_\mathrm{e}\ $}
\newcommand{\lm}{$\lambda$}

\title[Kinematical analysis of Hf 2-2 and M 1-42]{Kinematical analysis of PNe with high ADF: Hf 2-2 and M 1-42}

\author[Castañeda-Carlos et al.]{
L. C. Castañeda-Carlos, $^{1,3}$\thanks{E-mail: lcastaneda@astro.unam.mx}
M. G. Richer,$^{2}$
S. Torres-Peimbert,$^{1}$ 
A. Arrieta$^{3}$ and
L. Arias$^{3}$
\\
$^{1}$Universidad Nacional Autónoma de México. Instituto de Astronomía,  Apartado Postal 70-264, CP 04510 Ciudad de México, México\\
$^{2}$Universidad Nacional Autónoma de México, Instituto de Astronomía,  KM 107 Carretera Tijuana-Ensenada, CP 22860 Ensenada, Baja California, M\'exico\\
$^{3}$Universidad Iberoamericana, Departamento de Física y Matemáticas,  Prol. Paseo de la Reforma 880, Lomas de Santa Fe, CP 01210, Ciudad de México, México
}

\date{Accepted XXX. Received YYY; in original form ZZZ}

\pubyear{2024}

\begin{document}
\label{firstpage}
\pagerange{\pageref{firstpage}--\pageref{lastpage}}
\maketitle

\begin{abstract}

We use deep Echelle spectroscopy of the planetary nebulae Hf 2-2 and M1-42 to study the characteristics of the plasma that gives rise to their high abundance discrepancy factors (70 and 20, respectively).  We analyze position-velocity diagrams for forbidden and permitted lines (92 and 93 lines in Hf 2-2 and M 1-42, respectively), to compare their kinematic behaviour and to determine the physical characteristics of the emitting plasma.  We confirm that there are two plasma components in both nebulae: a normal nebular plasma that emits both forbidden and permitted lines and an additional plasma component that emits the permitted lines of \ion{O}{i}, \ion{C}{ii}, \ion{N}{ii}, \ion{O}{ii}, and \ion{Ne}{ii}.  These plasma components have different spatial distributions, with the additional plasma component being the more centrally concentrated.  Their physical conditions are also different, with the additional plasma component being denser and cooler.  We find that, in these objects, the additional plasma component contains masses of N$^{2}$ and O$^{2}$ ions that are at least as large as the normal nebular plasma.  In both objects, we find strong gradients in the electron temperature in small volumes near the central star.  Compared to NGC 6153, we find that the larger ADFs in Hf 2-2 and M 1-42 are due to larger masses of ions that emit only in the permitted lines, and not due to the physical conditions.

\end{abstract}




\section{Introduction}

Since 1942 \citep{wyse1942ApJ}, it has been known that the abundance ratios in ionized gas measured from permitted lines are almost always higher than those measured from forbidden lines. This systematic difference is known as the abundance discrepancy and has been extensively studied since the 1980s (when CCDs came into general use) in star-forming regions (\ion{H}{ii} regions) and remnants of evolved stars, such as planetary nebulae (PNe). Several hypotheses have been proposed to explain this difference, with the best studied hypotheses being temperature inhomogeneities \citep{1967ApJPeimbert} and chemical composition inhomogeneities or multiple plasma components \citep{torrespeimbert1990, liu2000ngc6153}.  \citet{tsamis2004} introduced the well-known abundance discrepancy factor (ADF) to characterize the discrepancy between the abundance ratios derived from permitted and forbidden lines. 

The ADF measured in \ion{H}{ii} regions has a median value of 1.9, while in planetary nebulae the median value is slightly higher at 2.6 \citep{wesson2018}.
\footnote{See \href{https://nebulousresearch.org/adfs/}{https://nebulousresearch.org/adfs/} for the most up-to-date values.}  However, the value for planetary nebulae includes the $\sim 20-25\%$ of PNe with very high ADF values ($>5$), among which Abell 46 has the highest ADF of 120 \citep[][we exclude the born-again PN Abell 30]{Corradi_2015}. Recently, \citet{mendez-delgado2023Natur} analyzed the ADF in a large sample of ionized nebulae. They found that, in \ion{H}{ii} regions, the discrepancy can be explained by temperature inhomogeneities in the most highly ionized gas.  However, for planetary nebulae, these inhomogeneities cannot fully account for the high ADF values.

Using high-spectral-resolution spectroscopy, \citet[][\citeyear{richer2022ngc}]{richer2013ApJ} studied the spatially-resolved kinematics of the nebular shell in NGC 7009 and NGC 6153. In both objects, the kinematics of the emission lines imply the apparent presence of two plasma components, a normal nebular plasma that emits in all emission lines, and an additional plasma component that emits only in permitted lines of \ion{C}{ii}, \ion{N}{ii}, \ion{O}{ii}, and \ion{Ne}{ii} (and \ion{O}{i} in NGC 6153).  The plasma component responsible for the permitted lines of \ion{C}{iii}, \ion{N}{iii}, and \ion{O}{iii} is unclear.  Likewise, the physical conditions differ in these two plasma components, with the additional plasma component being denser and colder.  In both objects, the additional plasma component likely accounts for about half the mass of $\mathrm O^{2+}$ ions (and $\mathrm N^{2+}$ ions in NGC 6153).  In NGC 6153, they were able to demonstrate that the additional plasma component is deficient in H.  Also in NGC 6153, they found that the normal nebular plasma contains temperature fluctuations whose amplitude varies spatially.  
\citet{pena2017}, also using high spectral resolution spectroscopy, measured the expansion velocities of forbidden and permitted lines in 14 PNe. They found that the permitted lines exhibit lower expansion velocities than the forbidden lines for a given ion, concluding that the component emitting the permitted lines is located closer to the central star.  

The evidence in the literature that supports the coexistence of two components with different physical conditions in PNe is not limited to high-spectral-resolution spectroscopy.  For instance, \citet{zhang2005} found that the temperature derived from \ion{He}{i} lines is lower than that determined from the Balmer decrement in \ion{H}{i}, which is inconsistent with a temperature fluctuation scenario.  \citet{Storey-and-sochi2014MNRAS} studied the continuous spectrum of Hf 2-2 and found that the best model reproducing the observations requires two components with significantly different temperatures.  
More recently, \citet{garciarojas2022} and \citet{gomezllanosetal24}, using medium-resolution, integral field spectroscopy, found strong spatial variations in the ADF in Hf 2-2, M 1-24, NGC 6153, and NGC 6778.

Here, we consider high-spectral-resolution spectroscopy to study the physical conditions and chemical abundances in Hf 2-2 and M 1-42, both of which have high ADFs.  Hf 2-2 is a round planetary nebula with a markedly high ADF of 70–80 \citep{liu2006,mccnabb2016}.  \citet{1998lutz} and \citet{2000bond} identified the central star of this nebula as a binary system with a period of 0.4 days.  
Based upon direct imaging, M 1-42 is a bipolar planetary nebula \citep{guerreroetal2013,tan2023} whose ADF is also very high, $20-22$ \citep{liuetal2001,mccnabb2016}.  There is no information regarding the binarity of its central star. 
\citet{garciarojas2022} find lower, though still high, ADFs with values of 7.9 and 18.2 in M 1-42 and Hf 2-2, respectively.  They also concluded that the two plasma components in both objects contain similar amounts of oxygen.  

We present a comprehensive study of the kinematics in Hf 2-2 and M 1-42 from the high-resolution, long slit observations available in the European Southern Observatory (ESO) data archive.  In Section 2 we describe the observations, our data reduction, and the reddening correction we derive.  In Sections 3 and 4, we describe for Hf 2-2 and M 1-42, respectively, our results concerning the kinematics, physical conditions, ionic abundances, ADF, and the relative N$^{2+}$ and O$^{2+}$ masses in the two apparent plasma components.   We discuss these results in Section 5 and present our final conclusions in Section 6.

\section{Observations, data reduction and PV diagrams}

This section presents the observations of Hf 2-2 and M 1-42, along with the corresponding data reduction procedures. Subsequently, we describe the construction of position-velocity (PV) diagrams used to analyze the kinematics, physical conditions, and elemental abundances. The section concludes with the determination of interstellar reddening.  These observations were conducted as part of the same program and during the same night as those for NGC 6153, so we provide only a brief summary of the observations and their reduction and refer the reader to \citet{richer2022ngc} for further details.

\subsection{Observations and data reduction}
\label{sec:obs} 

\begin{figure}
	\centering
	\includegraphics[width=\linewidth]{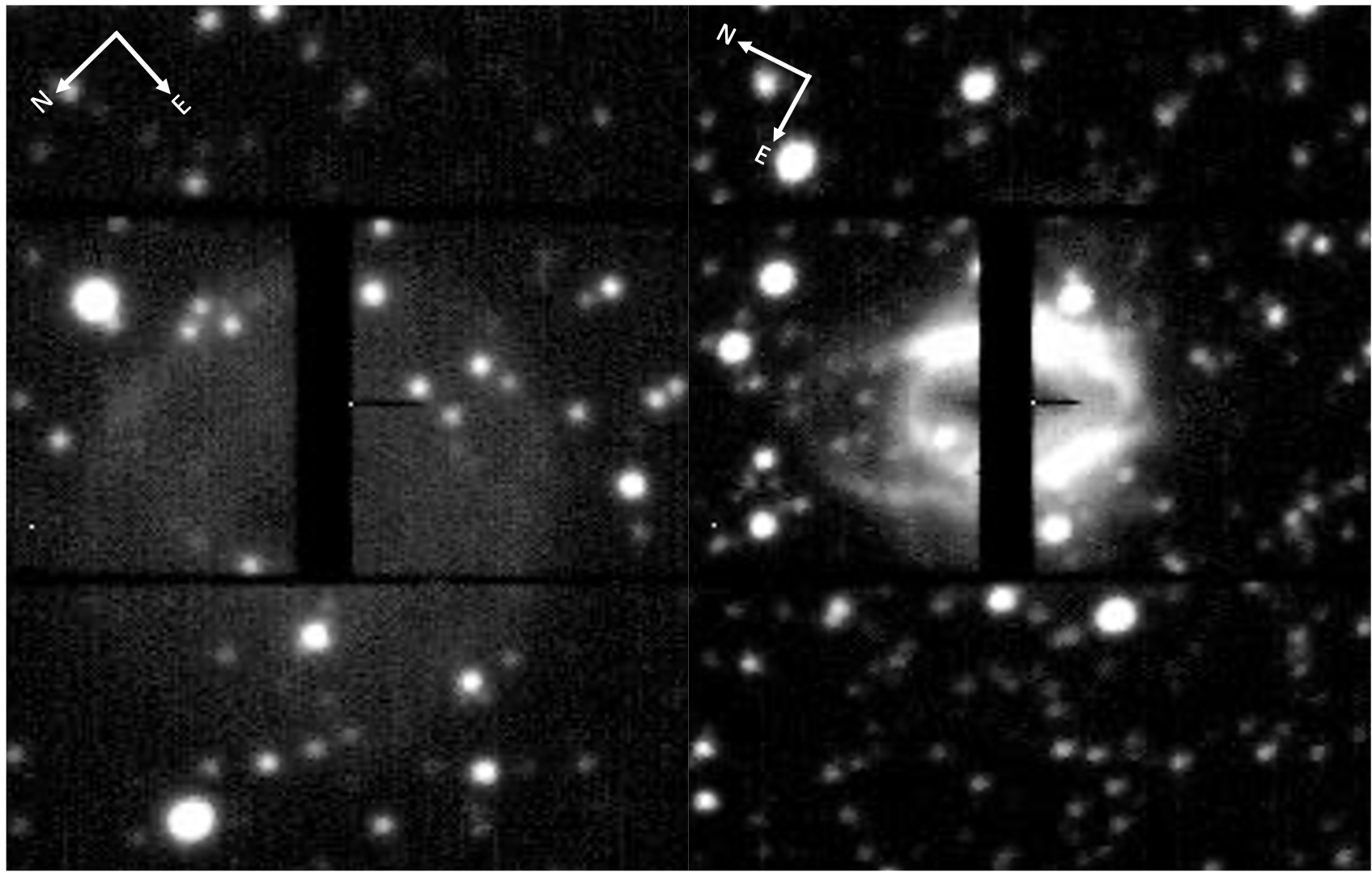}
	\caption{These images of Hf 2-2 (left) and M 1-42 (right) indicate the slit position for the red spectra. The slit size is $15.75\arcsec \times 2\arcsec$  but the decker limits the silt length to $13\arcsec$ for red spectra and 10\arcsec for blue. The arrows show the image orientation.  The slit position angle is $45^\circ$ for Hf 2-2 and $120^\circ$ for M 1-42.}  
	\label{fig:slits}
\end{figure}

The data were retrieved from the ESO data archive.  The spectra were obtained on 2002 June 8 as part of program ID 69.D-0174(A) using the Ultraviolet and Visual Echelle Spectrograph (UVES, \citealp{dekker2000}) on the Very Large Telescope (VLT) of the ESO. 
The observations were taken at four long-slit wavelength intervals with two instrument configurations. Each configuration comprises three parameters, one of the dichroics (DIC1/DIC2), a cross-disperser on each arm and an angle for the cross-disperser. Figure \ref{fig:slits} shows the location of red slit in both objects.
For both Hf 2-2 and M 1-42, two 1800s exposures were obtained for all wavelength intervals. In addition, for M 1-42,  a single short exposures of 60s for all the intervals were obtained to avoid saturating the brightest lines.
The blue and red slits are 10\arcsec and 13\arcsec long with a spatial resolution per pixel of 0.5\arcsec and 0.36\arcsec, respectively. 

Data reduction was performed using the Image Reduction and Analysis Facility (IRAF; \citealp{Tody1986}, \citeyear{Tody1993}; \citealp[]{Fitzpatricketal2024}). Since the observations are part of the same program and were carried out on the same night as those of NGC 6153, we used the dispersion and sensitivity functions determined for that object, which are described in detail in \citet{richer2022ngc}.

\subsection{Construction of PV diagrams}

Position-velocity maps are a tool used to study the distribution and kinematics of gas at different spatial positions within extended nebulae. The map is constructed by plotting the velocity of emission lines as a function of spatial position along the nebula using a long slit with high spectral resolution.

The PV diagrams were created using Python routines. The two-dimensional spectra were sliced in the spatial direction into sections 1 pixel high.  The slices were interpolated to build position-velocity maps with constant spatial and velocity scales of 0.36\arcsec/pixel and 1\kms/pixel, respectively. The zero position in the spatial direction was set at the location of the central star.  The velocity was determined using the Doppler effect, with zero velocity set at the systemic velocity, based on the laboratory wavelength of each emission line.  

\begin{figure}
    \centering
    \includegraphics[width=\linewidth]{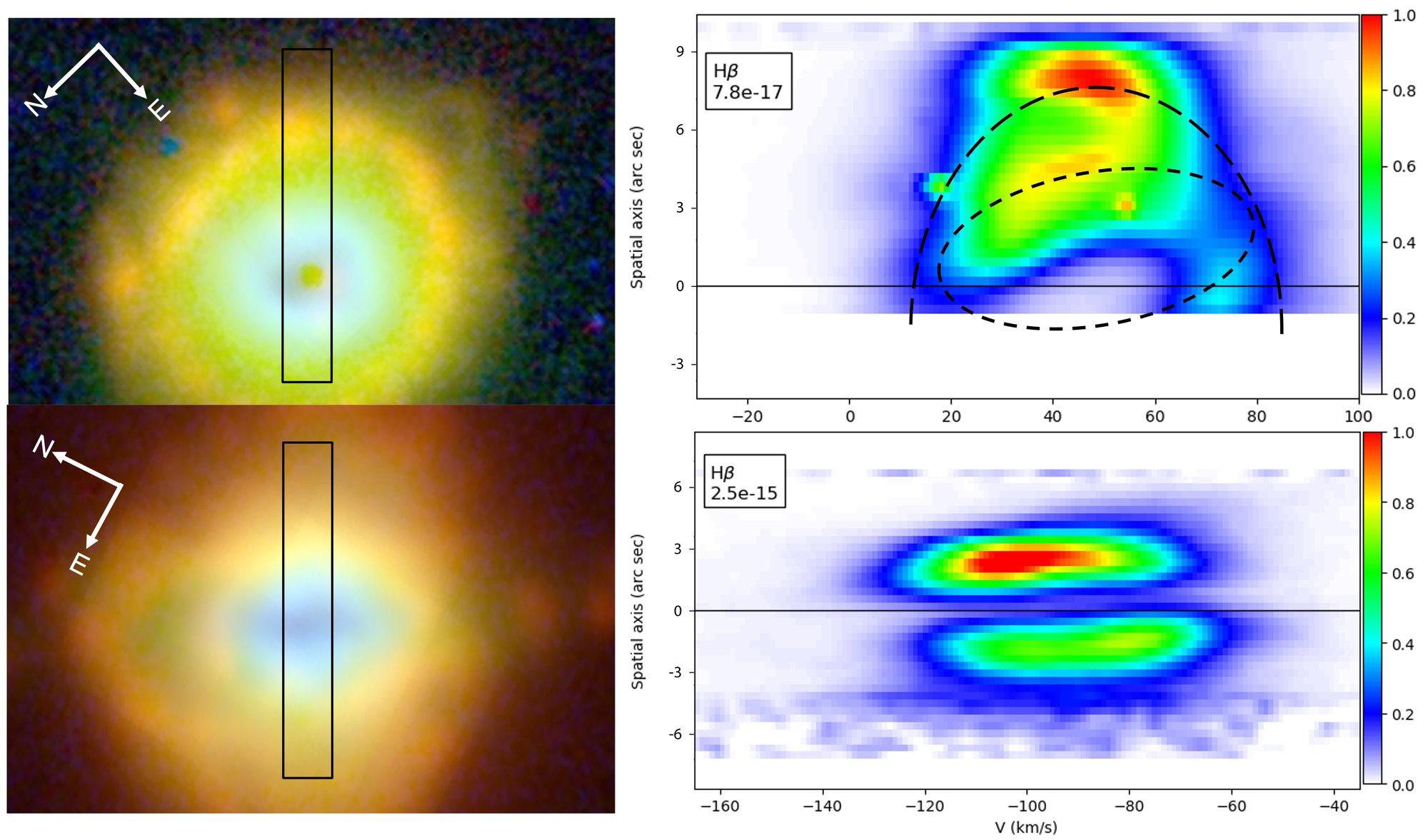}
    \caption{Left:  We present RGB composite images based upon MUSE data cubes for Hf 2-2 (top) and M 1-42 (bottom) with the spectrograph slit superposed \citep{garciarojas2022}. Emission from \ion{He}{ii} $\lambda$4686 is shown in blue, [\ion{O}{iii}] $\lambda$4959 in green, and [\ion{N}{ii}] $\lambda$6548 in red, all displayed on a logarithmic scale.  Right:  We present the PV diagrams of the H$\beta$ line for each object at the same spatial scale as the images on the left.  Here, and in subsequent PV diagrams, the line intensity runs from zero (white) to its maximum value (red), given in the box at upper left.  The vertical (spatial) scale is in arc seconds, with a resolution of $0.36 \arcsec/$pixel, and the horizontal scale is the velocity in \kms, centred at the systemic velocity of each object.  The horizontal black line indicates the position of the central star.  The blank space at the bottom of the PV diagram for Hf 2-2 is the spatial region seen only in the red slit.  The PV diagrams have the nebular and stellar continua subtracted.  Two schematic ellipses have been drawn to describe the inner and outer shells in Hf 2-2.}
    \label{fig:PV}
\end{figure}

Figure \ref{fig:PV} shows the PV diagrams for the H$\beta$ line of the two planetary nebulae. On the left side, the position of the red slit is overlaid on an image of the nebula taken from the data cubes of \citet{garciarojas2022}. On the right side, the profiles show the line decomposition caused by gas motion along the slit.
In the top right of Figure \ref{fig:PV}, the PV diagram of Hf 2-2 shows a double-shell morphology, indicated by two ellipses in a dashed line. The inner shell extends to $\sim$ 4.5\arcsec\ and the outer shell to $\sim$7.5\arcsec\ from the central star. The star’s position is marked by the horizontal line. The maximum velocity difference for the outer shell (the difference between the approaching and receding components) is about 70\kms. In the bottom panel of the same figure, M 1-42 shows a slightly more compact spatial emission. The position of the star is also indicated by a horizontal line.
In these diagrams we have removed the cosmic rays that do not fall on the emission from the objects.  Removing these cosmic rays does not affect our results.

We often need to compare PV diagrams for several emission lines.  To avoid introducing spurious artefacts, the PV diagrams must be aligned precisely in both spatial and velocity coordinates.  Given the faintness of the central stars, their positions can be uncertain by up to $\pm 1$\,pixel.  Likewise, between the errors in our wavelength calibration and the uncertainties in laboratory wavelengths, it is often necessary to apply shifts of up to several \kms\ to align them.  In these cases, we set the systemic velocity for all lines involved to the same value (given below for each object).  

\subsection{Interstellar reddening}\label{sec_reddening}

Determining and correcting for interstellar extinction is crucial to our analysis of the physical conditions and the chemical abundances in Hf 2-2 and M 1-42.  Since we use the same method implemented by \citet{richer2022ngc}, the reddening correction is also necessary to place the six spectral intervals on a common flux scale.  
To compute the reddening, we sum the two-dimensional spectrum over the spatial extent of the blue slit, thereby simulating a one-dimensional spectrum. We adopted the \citet{fitzpatrick1999correcting}  reddening law, after scaling to a total-to-selective extinction ratio of $R_{V} = 3.07$ \citep{2000mccall}.  We used the intensity of the \ion{H}{i} and \ion{He}{i} lines with respect to H$\beta$ 
and \ion{He}{i} $\lambda\lambda$4922,5876, respectively, and the atomic data cited in Table \ref{tab:atomic-data}.  We used the average of the emissivities at electron densities of 1,000 and 10,000 cm$^{-3}$ calculated for an electron temperature of 10,000 K, typical conditions for planetary nebulae.  \citet{richer2022ngc} provide further details.

Our derived reddening values for Hf 2-2 and M 1-42 are $E(B-V) = 0.0\pm 0.15$ mag and $0.35\pm 0.13$ mag, respectively, equivalent to $c(\mathrm H\beta)=1.424 E(B-V) = 0.0\pm 0.21$ and $0.50\pm 0.19$ dex, respectively.  Given that we adopt the flux calibration used by \citet{richer2022ngc} (see \S \ref{sec:obs}), our extinction values for Hf 2-2 and M 1-42 do not agree with other values in the literature, but do yield the correct flux scale and relative intensities over all six wavelength intervals.

\section{H\MakeLowercase{f} 2-2}
\label{sec:hf22}

We begin with the analysis of Hf 2-2. In this section, we start with the ionization structure, followed by the kinematic analysis, and then proceed to the electron temperature and density diagnostics to estimate the abundances and calculate the ADF.

\begin{table}
\centering
\caption{Atomic data used}
\begin{tabular}{cl}
\hline
Ion    & \multicolumn{1}{c}{Atomic data}                     \\ \hline
\ion{H}{i}    & \cite{storey1995recombination} 
\\
\ion{He}{i}   & \cite{porter2013}              \\
\ion{He}{ii}  & \cite{storey1995recombination}
\\
\begin{tabular}[c]{@{}l@{}} \ion{N}{ii} \\ \\ \end{tabular} &
  \begin{tabular}[c]{@{}l@{}}\cite{Nussbaumer1984}; \cite{Pequignot1991}; \\ \cite{Fang2011}; \cite{Tayal2011}\end{tabular} \\
\begin{tabular}[c]{@{}l@{}} \ion{O}{ii} \\ \\ \\ \end{tabular} &
  \begin{tabular}[c]{@{}l@{}}\cite{Zeippen1982}; \cite{Nussbaumer1984}; \\ \cite{Wenaker1990};  \cite{Pequignot1991}; \cite{Wiese1996}; \\ \cite {Tayal2007};  \cite{Kisielius2009}; \cite{Storey2017}\end{tabular} \\
\begin{tabular}[c]{@{}l@{}} \ion{O}{iii} \\ \\ \\ \\ \end{tabular} &
  \begin{tabular}[c]{@{}l@{}}\cite{Nussbaumer1984}; \cite{Pequignot1991}; \\ 
  \cite{Storey2000}; \cite{Tachiev2001}; \\ \cite{Froese2004}; \cite{Storey2014rec}; \\ \cite{Kramida2021}\end{tabular} \\
\ion{S}{ii}   & \cite{Rynkun2019}; \cite{Tayal2010}                 \\
\hline
\end{tabular}
\label{tab:atomic-data}
\end{table}

\subsection{Kinematical analysis} \label{sub:kinehf22}

\begin{figure}
    \centering
    \includegraphics[width=\linewidth]{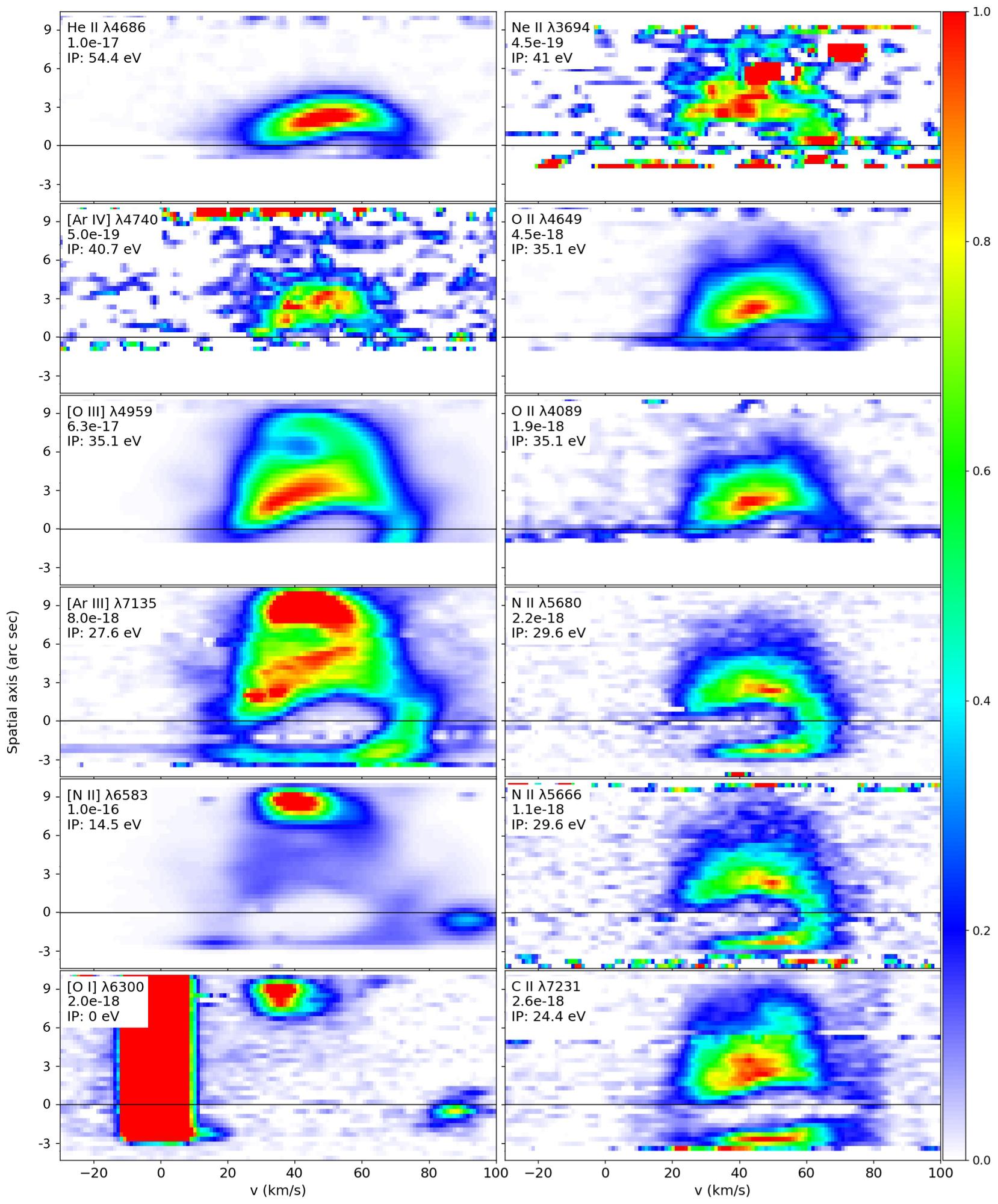}
    \caption{These PV diagrams illustrate the ionization structure of Hf 2-2. The left column shows commonly observed strong forbidden lines as well \ion{He}{ii} $\lambda$4686. The right column displays weak permitted lines of C$^+$, N$^+$, and O$^+$. In [\ion{O}{i}] $\lambda$6300, the bright vertical band is a sky line. The PV diagrams are arranged in order of ionization potential, from lowest (bottom) to highest (top). Here, the ionization energy corresponding to the ion that initiates the line emission process is shown in the box at upper left.}
    \label{fig:ion}
\end{figure}

In order to study the stratification of gas in Hf 2-2, Figure \ref{fig:ion} presents a selection of PV diagrams of emission lines produced by ions with different degrees of ionization.  The velocity scale is centred on the systemic velocity of the nebula that we measure, 54\,\kms\ (heliocentric).  On the left, we present forbidden lines, as well \ion{He}{ii} $\lambda$4686. On the right, we display some permitted lines of \ion{C}{ii}, \ion{N}{ii}, \ion{O}{ii}, and \ion{Ne}{ii}. For both columns, the profiles have been arranged from the highest to lowest lowest ionization potential. For the forbidden lines, the ionization potential ranges from 0 eV (for [\ion{O}{i}] $\lambda$6300) to 54.4 eV (for \ion{He}{ii} $\lambda$4686). For the permitted lines shown, the ionization potential ranges from 24.4 eV to 41 eV, i.e., the range spanned from [\ion{Ar}{iii}] to [\ion{Ar}{iv}].

A systematic change in the morphology of the PV maps is observed in the left-hand column of Figure \ref{fig:ion} as a function of the degree of ionization. At the highest ionization, such as \ion{He}{ii} $\lambda$4686 and [\ion{Ar}{iv}] $\lambda$4740, only the inner shell is discernible.  The outer shell becomes visible in the [\ion{O}{iii}] $\lambda$4959 line,  and increases in brightness relative to the inner shell in lower ionization lines, namely [\ion{Ar}{iii}] $
\lambda$7135 and [\ion{N}{ii}] $\lambda$6583.  Ultimately, the inner shell is no longer visible in [\ion{O}{i}] $\lambda$6300.  This behavior is consistent with nebular physics, where the degree of ionization decreases with distance from the central star.

In contrast, the PV diagrams of the permitted lines in the right column of Figure \ref{fig:ion} are similar at all ionization potentials.  This suggests that the emission originates from the same volume of the nebula regardless of the degree of ionization. The region emitting the permitted lines is apparently located in the inner shell.  This is especially striking considering that the ionization potentials in the right column span those for the PV diagrams dominated by emission from the inner shell ([\ion{Ar}{iv}] $\lambda$4740) to those dominated by the outer shell ([\ion{Ar}{iii}] $\lambda$7135) in the left column. 

In Figure \ref{fig:ion}, there is a noticeable change in the velocity extent of the inner shell as a function of the degree of ionization.  That is, the velocity extent is greater for low ionization lines. To quantify this, we measured the velocity splitting of the inner shell for different lines. Specifically, we fitted Gaussian profiles to the components of the line corresponding to the receding and approaching sides of the shell. These velocities were measured over of 3 pixels (1.08\arcsec) at a position centred 1.44\arcsec\ to the SW of the central star (4 rows above in the PV diagrams).  The difference in the velocities of the components allows us to estimate the velocity splitting of the gas as a function of the degree of ionization.  Velocities were measured for 90 lines. The lines were identified according to the ion and the different excitation processes responsible for the emission. When more than one line was available, the average velocity splitting and the standard deviation were calculated.

Table \ref{tab:hf2} presents the results of these measurements.  The first column indicates the ion that initiates the transitions.  For recombination lines, this is the ion that recombines with an electron.  The second column shows the energy required to generate that ion.  Usually, this is the ionization energy for the ion that emits the line, e.g., for \ion{O}{ii} lines, the O$^{2+}$ ion initiates the transition, recombining with an electron, so the relevant energy is the ionization potential of O$^+$.  We adopt ionization energies from \citet{Kramida2021}.  The Bowen fluorescence lines of \ion{N}{iii} and \ion{O}{iii}, however, are plotted at the ionization energy of He$^+$ because they are initiated by the Ly$\alpha$ transition of \ion{He}{ii}, due to the recombination of He$^{2+}$.  The third column lists the average velocity splitting and its standard deviation (when several lines could be measured).  The fourth column lists the emission lines used for the respective ion along with their excitation processes. 

\begin{table}
\centering
\caption{Ionization potential and velocity splitting of the observed ions in the central region of Hf 2-2}
\resizebox{0.48\textwidth}{!}{
\begin{tabular}{@{}cccl@{}}
\toprule
  Parent        & E$_{ion}$ & Velocity Splitting & \multicolumn{1}{c}{Lines (\AA)}                                                                   \\ 
                     &(eV)& (km~s$^{-1}$) & \\
\midrule
\begin{tabular}[c]{@{}l@{}} H$^{+}$ \\ \\\end{tabular}  & \begin{tabular}[c]{@{}l@{}} 13.6  \\ \\ \end{tabular}    & \begin{tabular}[c]{@{}l@{}} 41.0 $\pm$ 1.3     \\ \\ \end{tabular}           &  \begin{tabular}[c]{@{}l@{}}  \ion{H}{i} $\lambda \lambda$3770, 3797, 3835, 3970, 4340, 4861, \\6563, 9229, 9545, 10049 \end{tabular} \\   
\begin{tabular}[c]{@{}l@{}} He$^{+}$  \\ \\ \end{tabular} & \begin{tabular}[c]{@{}l@{}} 24.6     \\ \\ \end{tabular} & \begin{tabular}[c]{@{}l@{}} 41.9 $\pm$ 5.0       \\ \\ \end{tabular}        & \begin{tabular}[c]{@{}l@{}} \ion{He}{i} $\lambda \lambda$3187, 3614, 4026, 4387, 4471, 4713, \\ 4922, 5015, 5047, 5875, 6678, 7281 \end{tabular}\\
\begin{tabular}[c]{@{}l@{}}  He$^{2+}$  \\ \\  \end{tabular} & \begin{tabular}[c]{@{}l@{}} 54.4     \\ \\  \end{tabular} & \begin{tabular}[c]{@{}l@{}} 22.7 $\pm$ 3.1    \\ \\  \end{tabular}           & \begin{tabular}[c]{@{}l@{}} \ion{He}{ii} $\lambda \lambda$3203, 4199, 4541, 4686, 4859, 5411, \\ 6560, 8236     \end{tabular}                   \\
C$^{0}$   & 0        & 38.3                         & {[}\ion{C}{i}{]} $\lambda$9850 (nebular)                                                                      \\
C$^{2+}$  & 24.4     & 29.4 $\pm$ 2.4               & \ion{C}{ii} $\lambda \lambda$5342, 6151, 6461, 6578, 7231, 9903                                     \\
N$^{+}$   & 14.5     & 41.5 $\pm$ 0.1               & {[}\ion{N}{ii}{]} $\lambda \lambda$6548, 6583 (nebular)                                                       \\
N$^{+}$   & 14.5     & 27.8                         & {[}\ion{N}{ii}{]} $\lambda$5755 (auroral)                                                                     \\
\begin{tabular}[c]{@{}l@{}} N$^{2+}$  \\ \\ \end{tabular} & \begin{tabular}[c]{@{}l@{}}  29.6   \\ \\ \end{tabular}  & \begin{tabular}[c]{@{}l@{}} 25.4 $\pm$ 1.9              \\ \\ \end{tabular} & \begin{tabular}[c]{@{}l@{}} \ion{N}{ii} $\lambda \lambda$4035, 4041, 5666, 5676, 5680, 5686, \\ 5710 \end{tabular}                               \\
N$^{2+}$  & 54.4     & 23.7 $\pm$ 3.9               & {[}\ion{N}{iii}{]} $\lambda \lambda$4097, 4634, 4641 (Bowen fl.)                                                \\
N$^{3+}$  & 47.4     & 33.9                         & \ion{N}{iii} $\lambda$4379                                                                          \\
O$^{+}$   & 13.6     & 37.8 $\pm$ 1.8               & {[}\ion{O}{ii}{]} $\lambda \lambda$3726, 3729 (nebular)                                                       \\
O$^{+}$   & 13.6     & 29.1 $\pm$ 0.5               & {[}\ion{O}{ii}{]} $\lambda \lambda$7319, 7330 (auroral)                                                       \\
O$^{+}$   & 13.6     & 33.5 $\pm$ 0.1               & \ion{O}{i} $\lambda \lambda$7775, 9265                                                              \\
O$^{2+}$  & 35.1     & 34.9                         & {[}\ion{O}{iii}{]} $\lambda$4959 (nebular)                                                                    \\
O$^{2+}$  & 35.1     & 26.9                         & {[}\ion{O}{iii}{]} $\lambda$4363 (auroral)                                                                    \\
\begin{tabular}[c]{@{}l@{}} O$^{2+}$  \\ \\ \end{tabular} & \begin{tabular}[c]{@{}l@{}} 35.1     \\ \\ \end{tabular} & \begin{tabular}[c]{@{}l@{}} 26.8 $\pm$ 2.9     \\ \\ \end{tabular}          & \begin{tabular}[c]{@{}l@{}}\ion{O}{ii} $\lambda \lambda$4072, 4084, 4089, 4317, 4349, \\ 4366, 4639, 4649, 4662, 4676, 6501   \end{tabular}    \\
O$^{2+}$  & 54.4     & 25.6 $\pm$ 4.3               & {[}\ion{O}{iii}{]} $\lambda \lambda$3133, 3312, 3340, 3344 (Bowen fl.)                                          \\
Ne$^{2+}$ & 41       & 33.9 $\pm$ 0.6               & {[}\ion{Ne}{iii}{]} $\lambda \lambda$3869, 3967 (nebular)                                                     \\
Ne$^{2+}$ & 41       & 33.9                         & \ion{Ne}{ii}  $\lambda \lambda$3568, 3694                                                           \\
Si$^{2+}$ & 16.3     & 22.5 $\pm$ 1.0               & \ion{Si}{ii} $\lambda \lambda$5041, 6371                                                            \\
S$^{+}$   & 10.4     & 42,7 $\pm$ 2.4               & {[}\ion{S}{ii}{]} $\lambda \lambda$6716, 6731 (nebular)                                                       \\
S$^{2+}$  & 23.3     & 42.3                         & {[}\ion{S}{iii}{]} $\lambda$9531 (nebular)                                                                    \\
S$^{2+}$  & 23.3     & 42.2                         & {[}\ion{S}{iii}{]} $\lambda$6312 (auroral)                                                                    \\
Cl$^{2+}$ & 23.8     & 40.7 $\pm$ 2.8               & {[}\ion{Cl}{iii}{]} $\lambda \lambda$5517, 5537 (nebular)                                                     \\
Cl$^{3+}$ & 39.6     & 25.5 $\pm$ 6.4               & {[}\ion{Cl}{iv}{]} $\lambda \lambda$7530, 8045 (nebular)                                                      \\
Ar$^{2+}$ & 27.6     & 42.3 $\pm$ 1.3               & {[}\ion{Ar}{iii}{]} $\lambda \lambda$7135, 7751 (nebular)                                                     \\
Ar$^{3+}$ & 40.7     & 27.8 $\pm$ 1                 & {[}\ion{Ar}{iv}{]} $\lambda \lambda$4711, 4740 (nebular)                                                      \\ \bottomrule                                             
\end{tabular}
}
\label{tab:hf2}
\end{table}

\begin{figure}
    \centering
    \includegraphics[width=\linewidth]{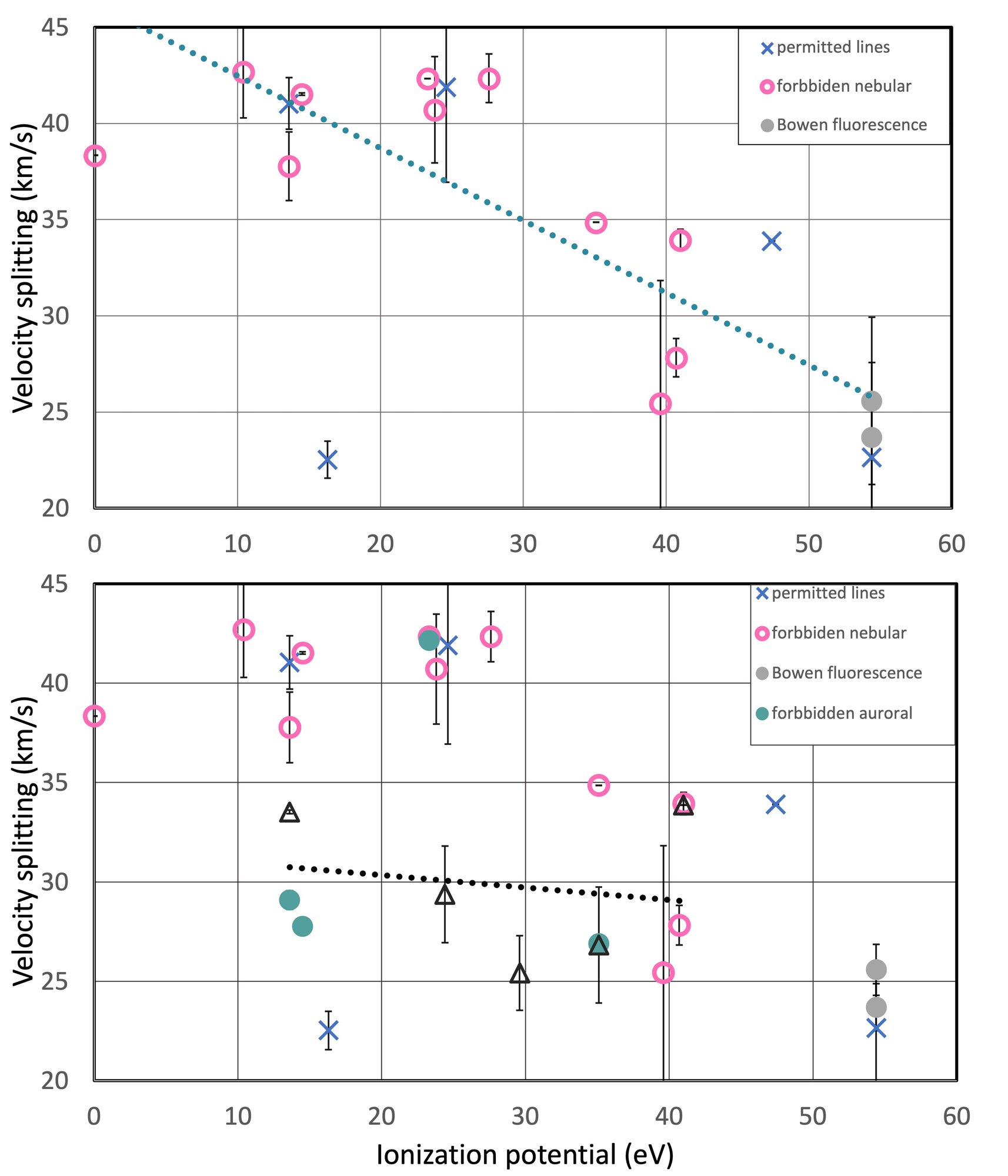}
    \caption{We present the Wilson diagram of Hf 2-2 in two parts. The upper diagram displays forbidden nebular, recombination, and fluorescence lines. The lower diagram repeats the results from the upper panel, but adds the data for forbidden auroral lines and the permitted lines of \ion{O}{i}, \ion{C}{ii}, \ion{N}{ii}, \ion{O}{ii}, and \ion{Ne}{ii}. The blue and black dotted lines represent linear fits to the collisionally excited lines and to the \ion{O}{i}, \ion{C}{ii}, \ion{N}{ii}, \ion{O}{ii}, and \ion{Ne}{ii} lines, respectively.}
    \label{fig:wd-hf22}
\end{figure}

Figure \ref{fig:wd-hf22} presents the results from Table \ref{tab:hf2}. The top panel shows the velocity splitting as a function of the ionization energy for forbidden nebular lines, recombination lines, and fluorescence lines.  A linear fit, excluding \ion{Si}{ii} (see below), reveals a negative slope, consistent with the findings of \citet{wilson1950ApJ}. This well-known result indicates that more highly ionized ions are located in an inner region of the nebula that expands more slowly.

In the bottom panel of Figure \ref{fig:wd-hf22}, we add the velocity splitting for the permitted lines of \ion{O}{i}, \ion{C}{ii}, \ion{O}{ii}, \ion{N}{ii} and \ion{Ne}{ii}. These lines exhibit a distinct kinematic behaviour, with a constant velocity splitting despite the considerable range of ionization energies required to produce the parent ions. Here, as in a growing number of cases \citep{richer2013ApJ,pena2017,richer2022ngc}, the behaviour of these lines defines what appears to be a distinct plasma component from that in the top panel of Figure \ref{fig:wd-hf22}.  

In the bottom panel of Figure \ref{fig:wd-hf22}, we also add the velocity splitting for the auroral forbidden lines.  Only [\ion{S}{iii}] $\lambda$6312 follows the trend found in the upper panel.  The others, [\ion{O}{ii}] $\lambda\lambda$7320,7330, [\ion{N}{ii}] $\lambda$5755, and [\ion{O}{iii}] $\lambda$4363 have substantially smaller velocity splitting than the nebular lines.  The first two are severely affected by a contribution due to recombination \citep{garciarojas2022}, which explains their anomalous kinematics.  The emissivity of the [\ion{O}{iii}] $\lambda$4363 line is undoubtedly affected by the very large temperature gradient (\S \ref{sub:forl}) and so is biased by the kinematics of the plasma closest to the central star. 

As for the \ion{Si}{ii} $\lambda\lambda$5041,6371 lines in the upper panel of Figure \ref{fig:wd-hf22} (Table \ref{tab:hf2}), \citet{grandi1976} argued that their emission originates from fluorescence excitation of the $4\mathrm d^{2}\,\mathrm D$ level \citep[$989-992$\AA][]{morton1991} due to starlight, as this excitation mechanism is several orders of magnitude more efficient than recombination.  These are likely minority Si$^{+}$ ions near the central star. 

The lines in the upper panel of Figure \ref{fig:wd-hf22} represent the majority of the ions and the majority of the mass in Hf 2-2.  Their kinematics adhere to a classical expansion law \citep[e.g.,][]{wilson1950ApJ}, namely a decrease in expansion rate with increasing ionization potential.  We denote this plasma as the \emph{normal nebular plasma} since it corresponds to the nebular plasma described in textbooks \citep[e.g.,][]{aller1984,osterbrockferland2006}.  In contrast, the \ion{O}{i}, \ion{C}{ii}, \ion{O}{ii}, \ion{N}{ii} and \ion{Ne}{ii} lines (like the auroral lines of [\ion{O}{ii}] and [\ion{N}{ii}]) present a distinctive kinematic behaviour.  Accordingly, we denote this plasma as the \emph{additional plasma component}. 

\subsection{Physical conditions} \label{sub:phycon}

This section describes the determination of the physical conditions. Diagnostics are performed for each pixel of the PV diagram. Before performing the calculation, each line is corrected for reddening by applying the correction factor $c$(H$\beta$) and using the reddening function of \citet{fitzpatrick1999correcting}. This is implemented using \texttt{PyNeb}, a package for the analysis of emission lines in gaseous nebulae, as developed by \citet{luridiana2014}. The \texttt{TemDen} module is employed to calculate the temperature and density from forbidden lines. For permitted lines, line ratios of \ion{N}{ii} and \ion{O}{ii} are used. A summary of the results is presented in Table \ref{summary}.

\subsubsection{Forbidden lines} \label{sub:forl}

\begin{figure*}
    \centering
    \includegraphics[width=0.89\linewidth]{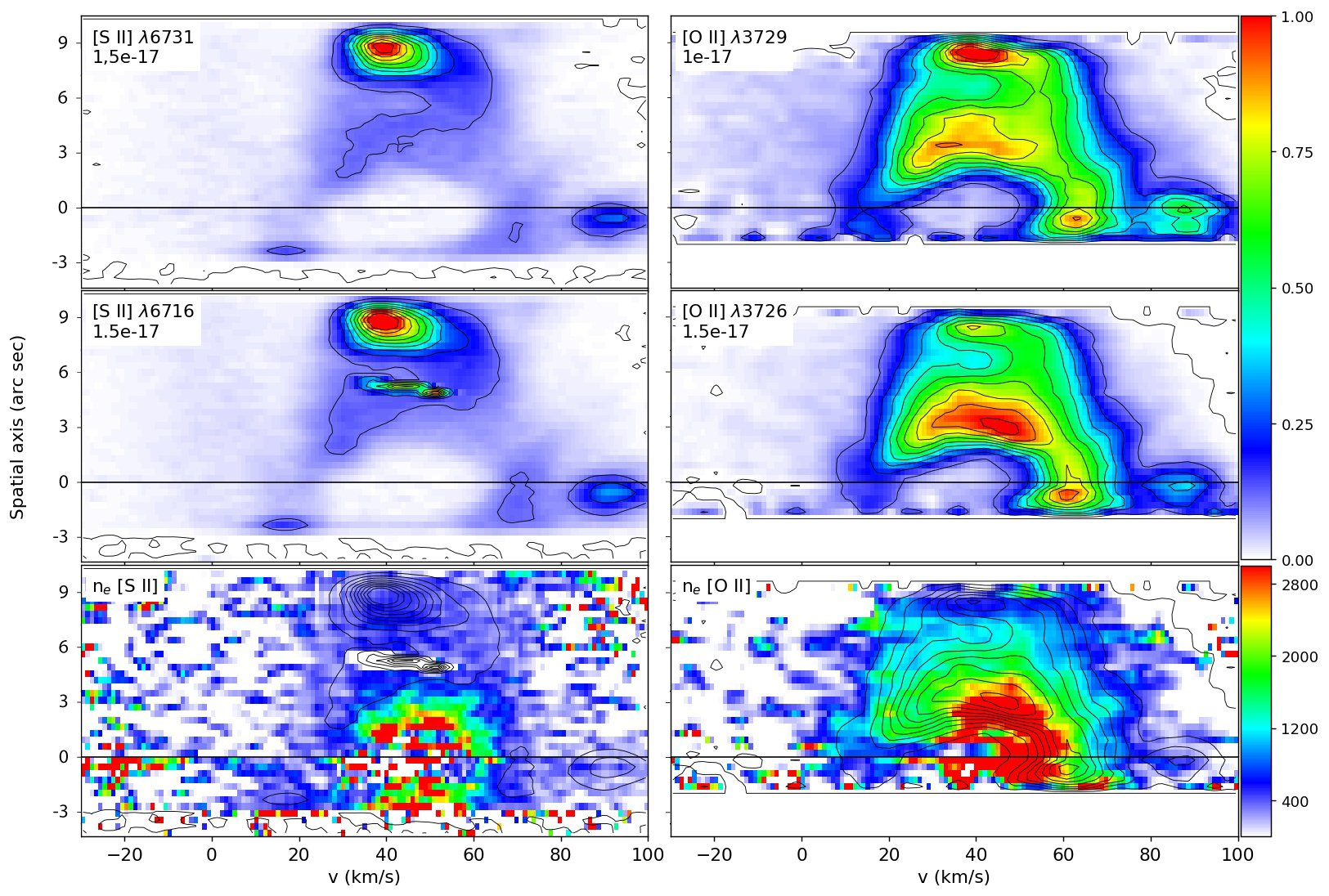}
    \caption{We present the PV diagrams of the [\ion{S}{ii}] $\lambda\lambda$6716,6731 (left) and [\ion{O}{ii}] $\lambda\lambda$3726,3729 lines (right) and the density derived from their ratios (bottom row) for Hf 2-2. The contour lines in the bottom row correspond to the intensities of [\ion{S}{ii}] $\lambda$6716 and [\ion{O}{ii}] $\lambda$3726 lines, respectively.  Here, and in subsequent PV diagrams, the contours are drawn at 10\%, 20\%, ..., 90\% of the maximum intensity, shown in the box at upper left.}
    \label{fig:den-oii-hf22}
\end{figure*}

To compute the electron density, $n_\mathrm{e}$, we use the [\ion{S}{ii}] $\lambda\lambda$6731/6716 and [\ion{O}{ii}] $\lambda\lambda$3726 and 3729 ratios.  The PV diagrams of the lines, as well as the density, are shown in Figure \ref{fig:den-oii-hf22}, assuming a \te = 10,000 K.

The PV diagrams of the [\ion{S}{ii}] $\lambda\lambda$ 6731,6716 lines, shown in the left column of Figure \ref{fig:den-oii-hf22}, reveal that the maximum emission is located in the outer shell of the nebula.  A high-velocity knot is also notable in both lines. In these regions, a low density of $500-1,000\,\mathrm{cm}^{-3}$ is found (bottom left panel). In the inner shell, where the emission from the [\ion{S}{ii}] $\lambda\lambda$ 6731,6716 lines is weak, higher densities of around 2,000\,cm$^{-3}$ are observed. In the innermost part of the inner shell, closest to the central star, the signal-to-noise ratio is too low and it is not possible to estimate the density (white pixels). Hence, we find a density contrast between the inner shell and the matter exterior to it, with \nel varying from $\sim 2,000$\,cm$^{-3}$ in the inner shell to to 500\,cm$^{-3}$ in the outermost gas, including the high-velocity knot.

We present the maps of the [\ion{O}{ii}] $\lambda\lambda$3729,3726 lines in Figure \ref{fig:den-oii-hf22}.  Although both [\ion{O}{ii}] and [\ion{S}{ii}] originate from low-ionization ions, the morphology of the emission observed in the two ions is distinct.  In [\ion{O}{ii}], the inner shell is much brighter than in [\ion{S}{ii}], suggesting the presence of a recombination component \citep[see also][]{liu2006}.  Even under these circumstances, the ratio of the [\ion{O}{ii}] $\lambda\lambda$3729,3726 lines is a valid density indicator, though the density will be that of the plasma component that dominates the emissivity.  As found for [\ion{S}{ii}], the density of the matter outside the inner shell is low, $\sim 500-1,000\,\mathrm{cm}^{-3}$.  However, the density found in the inner shell is much higher, up to $\sim 3,000\ \mathrm{cm}^{-3}$.  The difference, as we shall show below, is likely the result of a recombination contribution to the [\ion{O}{ii}] lines that is not present in [\ion{S}{ii}].

\begin{figure*}
    \centering
    \includegraphics[width=0.89\linewidth]{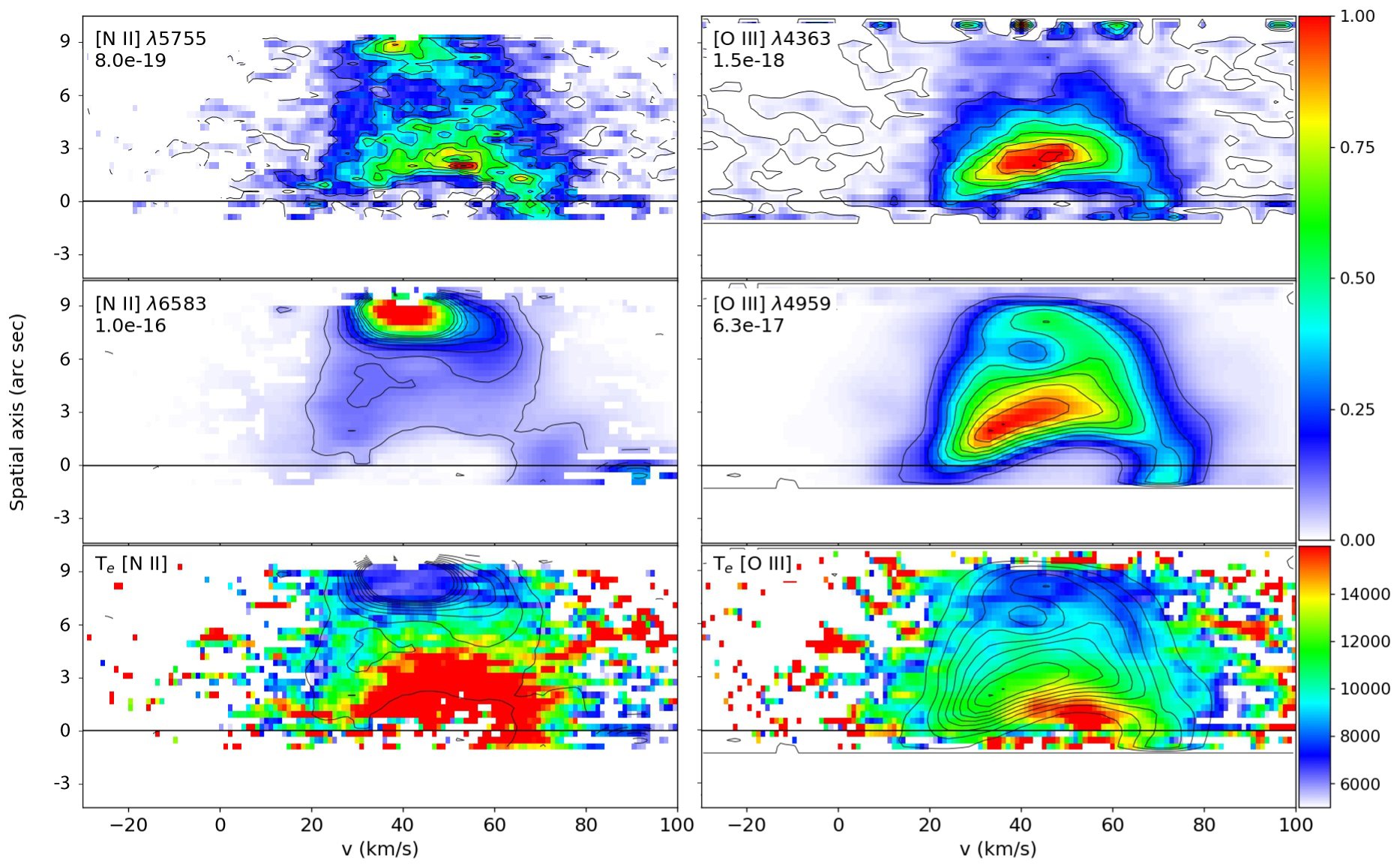}
    \caption{We present the PV diagrams of the [\ion{N}{ii}] $\lambda\lambda$5755,6583 (left) and [\ion{O}{iii}] $\lambda\lambda$4363,4959 lines (right) and the temperatures derived from them (bottom row) for  Hf 2-2 . We attempted to correct the [\ion{N}{ii}] lines to remove the contribution due to recombination (see text), but it was only partially successful.  The contours in the bottom row are of the intensity of [\ion{N}{ii}]  $\lambda$6583 and [\ion{O}{iii}] $\lambda$4959.}
    \label{fig:tem_cel_hf22}
\end{figure*}

To determine the electron temperature, \te, we use the ratios of the [\ion{O}{iii}] $\lambda\lambda$ 4363/4959 and [\ion{N}{ii}] $\lambda\lambda$5755/6583 lines. The PV diagrams of the lines, as well as the temperature distribution, are shown in Figure \ref{fig:tem_cel_hf22} assuming a \nel = 1,000 cm$^{-3}$.

For the [\ion{O}{iii}] lines (right side of Figure \ref{fig:tem_cel_hf22}), the inner shell is the brighter component, consistent with the trend illustrated in Figure \ref{fig:ion}.  We note that the contrast between the inner and outer shells is greater for the [\ion{O}{iii}] $\lambda$4363 line.  The bottom right panel of Figure \ref{fig:tem_cel_hf22} presents the map of the electron temperature derived from the [\ion{O}{iii}] $\lambda\lambda$4363/4959 line ratio.  The temperature map shows a clear gradient with temperatures below 8,000\,K in the outer shell, but rising to 10,000\,K in the inner shell and with values as high as 16,000\,K in a small volume close to the line of sight towards the central star.  

Since recombination and charge exchange can contribute to the emission in the [\ion{O}{iii}] lines, we now evaluate their relevance.  We use the \ion{O}{iii} $\lambda\lambda$3265,5592 lines, respectively, as probes of these processes and present their PV diagrams in Figure \ref{fig:reclines}.    Recombination can be ignored because the \ion{O}{iii} $\lambda$3265 line is absent and its intensity should be 3.6 times that of the recombination component of the [\ion{O}{iii}] $\lambda$4363 line \citep{richer2022ngc}.  Likewise, the \ion{O}{iii} $\lambda$5592 line is absent even though its intensity is expected to be twice that of the emission from [\ion{O}{iii}] $\lambda$4363 due to charge exchange \citep{richer2022ngc}, so this process is also irrelevant for the emission of [\ion{O}{iii}] $\lambda$4363 in Hf 2-2.  This analysis agrees with that of \citet{liu2006}.  (These processes are much less important for the excitation of [\ion{O}{iii}] $\lambda$4959 since collisional excitation of this line is easier than for [\ion{O}{iii}] $\lambda$4363.)  Hence, the strong temperature gradient in Figure \ref{fig:tem_cel_hf22} (bottom right) appears to be real.  

The velocity measured for the [\ion{O}{iii}] $\lambda$4959 line also supports the reality of the temperature gradient in Figure \ref{fig:tem_cel_hf22}.  The O$^{2+}$ ion should largely coincide with He$^+$ (and H$^+$), but we measure a substantially smaller velocity splitting for [\ion{O}{iii}] $\lambda$4959 (\S  \ref{sub:kinehf22}) than the \ion{H}{i} or \ion{He}{i} lines.  This is understandable if the emission from [\ion{O}{iii}] $\lambda$4959 is biased to the hotter inner regions of the O$^{2+}$ zone, which both expand more slowly and emit less intensely in the \ion{H}{i} and \ion{He}{i} recombination lines. 

The intensity of the [\ion{N}{ii}] $\lambda$5755 line in Hf 2-2 is known to include a strong contribution due to recombination \citep{liu2006,garciarojas2022}, so we attempt to ``decontaminate" this line following \citet{richer2022ngc}.  To do so, we need models of the emission from both plasma components.  We use the PV diagram of [\ion{O}{iii}] $\lambda$4959 to model the normal nebular plasma.  To model the emission from the additional plasma component, we decompose the \ion{N}{ii} $\lambda$5680 line and adopt the component attributed to the additional plasma component as our model \citep[details:][]{richer2022ngc}.  Figures  \ref{fig:hf22-nii-oii-decom} and \ref{fig:hf22-decon-nii} in the Appendix   show the intermediate steps.  We present the resulting PV diagrams for the [\ion{N}{ii}] $\lambda\lambda$5755,6583 lines in the top two panels of the left column in Figure \ref{fig:tem_cel_hf22}.  We present the electron temperature computed from the ratio of the [\ion{N}{ii}] lines in the bottom left panel of Figure \ref{fig:tem_cel_hf22}.  As found from the [\ion{O}{iii}] lines, there is a strong temperature gradient.  While the temperatures along the lines of sight $>6$\arcsec\ from the central star are similar in both [\ion{N}{ii}] and [\ion{O}{iii}], the [\ion{N}{ii}] lines indicate much higher temperatures close to the central star.  Likely, our prescription for ``decontaminating" the [\ion{N}{ii}] lines is insufficient, as we found for the case of NGC 6153 \citep{richer2022ngc}.  However, the [\ion{N}{ii}] temperature map confirms the lower temperatures from the [\ion{O}{iii}] lines in the outer part of Hf 2-2.  

Summarizing our results in this section, the outer shell in Hf 2-2 has a low electron density $500-1,000$\,cm$^{-3}$ and an electron temperature below 8,000\,K.  The inner shell is denser, with electron densities rising to at least 2,000\,cm$^{-3}$.  There appears to be a clear gradient in the electron temperature from below 8,000\,K in the outer shell to a typical value of 10,000\,K in the inner shell, but rising as high as 16,000\,K in a small volume near the central star.  The foregoing pertains to the normal nebular plasma.  If the additional plasma component dominates the emission from the [\ion{O}{ii}] $\lambda\lambda$3729,3726 lines towards the inner shell, its electron density is higher than found for the normal nebular plasma, at least 4,000\,cm$^{-3}$.

\subsubsection{Permitted lines} \label{sec_hf22_physcond_permitted}

To study the physical conditions for the additional plasma component, we use the permitted lines of \ion{N}{ii} and \ion{O}{ii} to calculate its electron temperature and density.  We follow the method described in \citet{richer2022ngc} using the atomic data from Table \ref{tab:atomic-data}.  

\begin{figure*}
    \centering
    \includegraphics[width=0.89\linewidth]{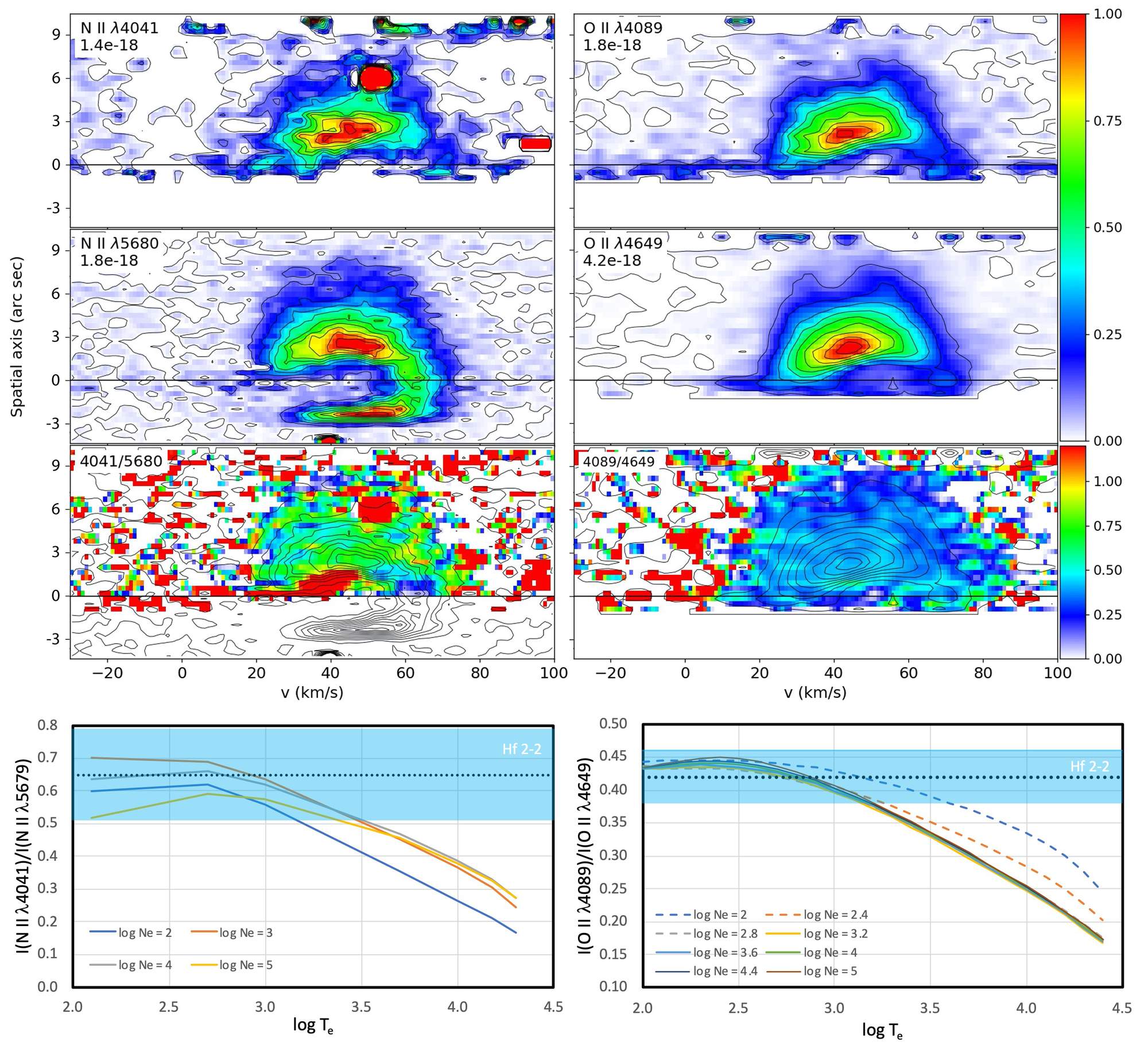}
    \caption{The two upper rows show the lines used in the calculation of \te using the \ion{N}{ii} (left) and \ion{O}{ii} lines (right) for Hf 2-2. The third row displays the line ratios with the isocontours of the intensity of the \ion{N}{ii} $\lambda$5680 and \ion{O}{ii} $\lambda$4649 lines overlaid, respectively. The bottom row presents the line ratio as a function of \te for different \nel values (Table \ref{tab:atomic-data}).  We indicate the mean value (dotted line) and standard deviation (blue shading), estimated from the PV diagrams of the line ratios \ion{N}{ii} $\lambda\lambda$4041/5680 and \ion{O}{ii} $\lambda\lambda$4089/4649 (third row).  A cosmic ray contaminates the PV diagram of \ion{N}{ii} $\lambda$4041, but is outside the area that we use in our analysis, defined by the 40\% intensity contour.}
    \label{fig:temrec-hf22-nii-oii}
\end{figure*}

Figure \ref{fig:temrec-hf22-nii-oii} shows the PV diagrams of the \ion{N}{ii} $\lambda\lambda$4041,5680 lines (first two panels). The third panel shows the ratio of these lines used to estimate the electron temperature.  The ratio map is relatively flat without significant gradients. Using pixels exceeding 40$\%$ of the maximum intensity in the PV diagram of $\lambda$5680, the mean \ion{N}{ii} $\lambda\lambda$4041/5680 ratio is $0.65\pm 0.14$. Finally, the bottom panel shows \ion{N}{ii} $\lambda\lambda$4041/5680 as a function of temperature for different \nel values. In this figure, the mean value we observe is represented by the horizontal dashed line, while the blue band indicates its uncertainty. The \ion{N}{ii} $\lambda\lambda$4041/5680 ratio restricts the electron temperature to values below 3,200 K.

We also estimate the electron temperature using the \ion{O}{ii} $\lambda\lambda$4089,4649 lines. In Figure \ref{fig:temrec-hf22-nii-oii}, the line profiles are shown in the top two panels of the figure and the \ion{O}{ii} $\lambda\lambda$4089/4649 line ratio in the third panel.  The line ratio shows a uniform mean value of $0.42\pm 0.05$ using all pixels exceeding 40\% of the maximum intensity in the PV diagram of the \ion{O}{ii} $\lambda$4649 line.  The bottom panel shows the line ratio as a function of \te for various \nel values. The ratio of the \ion{O}{ii} $\lambda\lambda$4089/4649 lines constrains \te to values below 2,000\,K, unless the density is very low, in which case \te $ < 3,200$\,K.

\begin{figure*}
    \centering
    \includegraphics[width=0.89\linewidth]{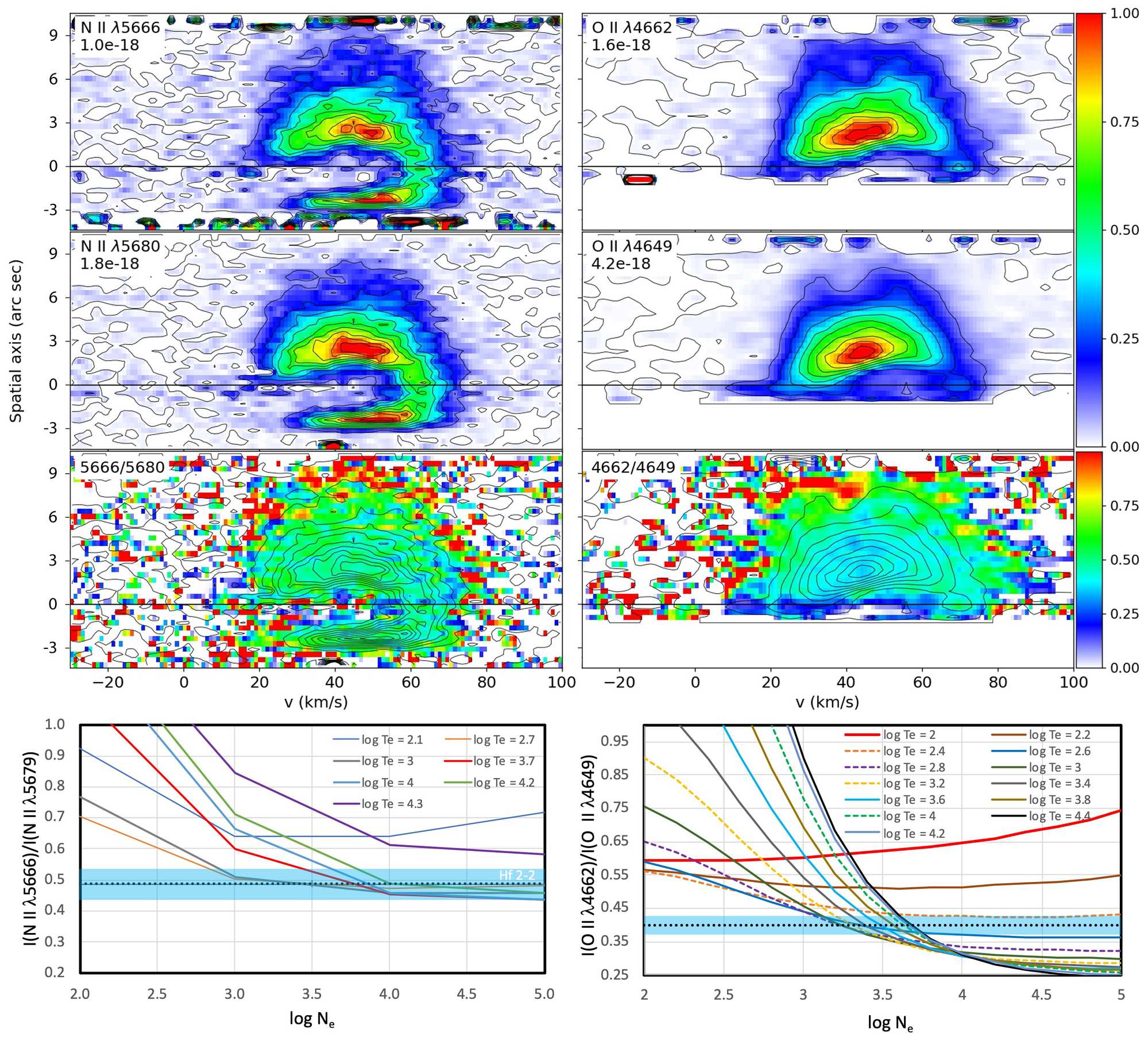}
    \caption{The top two rows show the lines used in the calculation of \nel from the  \ion{N}{ii} (left) and \ion{O}{ii} lines (right) for Hf 2-2. The third row displays the line ratios with the isocontours of the intensities of the \ion{N}{ii} $\lambda$5680 and \ion{O}{ii} $\lambda$4649 lines overlaid, respectively. The bottom row presents  the line ratio as a function of \te for different \nel values (Table \ref{tab:atomic-data}).  We indicate the mean value (dotted line) and standard deviation (blue shading), estimated from the PV diagrams of the \ion{N}{ii} $\lambda\lambda$5666/5680 and \ion{O}{ii} $\lambda\lambda$4662/4649 line ratios in the third row.}
    \label{fig:denrec-hf22-nii-oii}
\end{figure*}

In order to analyze the \nel in permitted lines, we employed the ratios of \ion{N}{ii} $\lambda\lambda$5666/5680  and \ion{O}{ii} $\lambda\lambda$4662/4649 (Figure \ref{fig:denrec-hf22-nii-oii}). The PV diagrams of the individual lines are shown in the first two panels of each figure, while the line ratios are displayed in the third panel. The plots of line ratios as a function of $n_\mathrm{e}$, shown in the bottom panel of each figure, constrain the range of electron density.  For the \ion{N}{ii} lines, the electron density exceeds 3,000\,cm$^{-3}$, unless the temperature is very low, in which case densities as low as 800\,cm$^{-3}$ are possible.  For the \ion{O}{ii} lines, the electron density is constrained to the $1,500-5,000$\,cm$^{-3}$ range, unless the electron temperature is in the $250-400$\,K range, in which case higher densities are allowed.  

Considering all of the foregoing restrictions simultaneously, the \ion{N}{ii} and \ion{O}{ii} lines imply an electron density of at least $3,000-5,000$\,cm$^{-3}$ and an electron temperature below 2,000\,K.  This electron density agrees with that based upon the [\ion{O}{ii}] $\lambda\lambda$3726,3729 lines if they are dominated by the recombination component in the inner part of inner shell (\S\ref{sub:forl}).  Hence, in Hf 2-2, the additional plasma component is clearly denser and colder than the normal nebular plasma.  

Table \ref{summary} summarizes the \te and \nel values derived from all indicators of the physical conditions.

\subsection{Ionic Abundances and ADF} \label{sec_Hf22_ADF}

We determine the ionic abundances of O$^{2+}$ for Hf 2-2 using the ratios of the [\ion{O}{iii}] $\lambda$4959/H$\beta$ and \ion{O}{ii} $\lambda$4649/H$\beta$ emission lines. Before calculating the ratios, we broaden the oxygen line profiles using a Gaussian profile to match the thermal width of H$\beta$ assuming temperatures of 2,000\,K for the \ion{O}{ii} $\lambda$4649 line and 8,000\,K for [\ion{O}{iii}] $\lambda$4959.

The $\mathrm O^{2+}/\mathrm H^+$ ionic abundances were calculated for each pixel in the PV diagrams using PyNeb's \texttt{getIonAbundance} function, ensuring consistency with the plasma conditions assigned to each component.  For the normal nebular plasma, we computed the $\mathrm O^{2+}/\mathrm H^+$ ionic abundance using the [\ion{O}{iii}] $\lambda$4959 line and adopting an electron density of 1,000 cm$^{-3}$ and the electron temperature from the PV diagram of the [\ion{O}{iii}] temperature (Figure \ref{fig:tem_cel_hf22}).  (The PV diagram of the [\ion{O}{iii}] temperature is first broadened to match the thermal width of H$\beta$ at a temperature of 8,000\,K.)  For the additional plasma component, we computed the $\mathrm O^{2+}/\mathrm H^+$ ionic abundance using the \ion{O}{ii} $\lambda$4649 line assuming an electron temperature of 2,000 K and an electron density of 5,000 cm$^{-3}$.  

\begin{figure}
    \centering
    \includegraphics[width=0.98\linewidth]{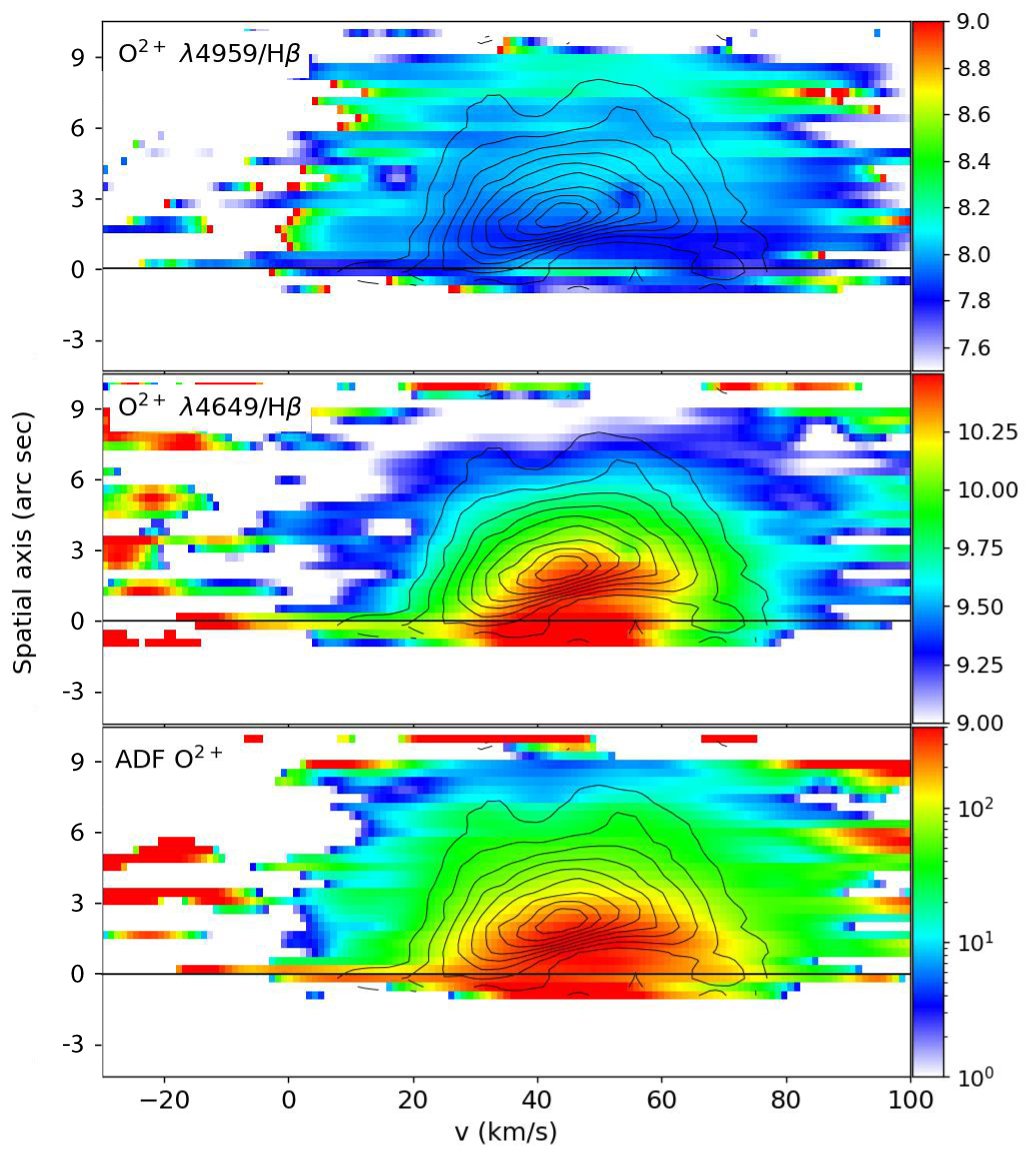}
    \caption{The upper two panels present the PV diagrams of the $\mathrm O^{2+}/\mathrm H^+$ ionic ratio in Hf 2-2 based upon the [\ion{O}{iii}] $\lambda$4959 (CEL, top) and \ion{O}{ii} $\lambda$4649 lines (RL, middle).  The bottom panel presents their ratio ($\mathrm{RL}/\mathrm{CEL}$), which is the ADF as defined by \citet{tsamis2004}.  Both abundance diagrams are plotted in units of $12+\log(\mathrm O^{2+}/\mathrm H^+)$, while the ADF diagram is displayed on a logarithmic scale. In all panels, the contours correspond to the intensity of the \ion{O}{ii} $\lambda$4649 line.}
    \label{fig:oiiabundances-adf-hf22}
\end{figure}

Figure \ref{fig:oiiabundances-adf-hf22} shows the PV maps of the O$^{2+}$ abundance based upon the two line ratios. 
While the $\mathrm O^{2+}/\mathrm H^+$ ionic abundance derived from the [\ion{O}{iii}] \lm4959 line is rather uniform, a strong abundance gradient is observed in the map derived from the \ion{O}{ii} \lm4649 line in the central region and at velocities close to systemic velocity. From the ratio of these two maps, we compute the PV diagram of the ADF for $\mathrm O^{2+}$. The resulting ADF map (bottom panel) generally follows that of the $\mathrm O^{2+}/\mathrm H^+$ ionic abundance derived from the \ion{O}{ii} \lm4649 line.  Along lines of sight close to the central star and velocities close to the systemic velocity in Hf 2-2, the ADF achieves values as high as 300.  Even at the extreme velocities (low and high) and in the outer shell, the ADF remains clearly in excess of 1.0.  

Our map of the ADF for $\mathrm O^{2+}$ in Hf 2-2 is congruent with that presented by \citet{garciarojas2022}.  In their map, the ADF varies from a factor of a few at the edge of the object to over a factor of 100 in the central regions.  We obtain a similar value of the ADF at the edge of the object, along the lines of sight farthest from the central star.  Towards the central regions, we obtain larger values than they observe, but that is expected since we resolve the variation in ADF along the line of sight.

\subsection{Relative masses of N$^{2}$ and O$^{2}$ in the plasma components}

We can use a decomposition of the \ion{N}{ii} $\lambda$5680 and \ion{O}{ii} $\lambda$4649 lines already described to account for the contribution of recombination in the [\ion{N}{ii}] lines to determine the mass fraction of N$^{2+}$ and O$^{2+}$ ions in the additional plasma component \citep[for details, see][Figure  \ref{fig:hf22-nii-oii-decom}]{richer2022ngc}.  In both cases, the total emission in these lines is strongly dominated by the emission from the additional plasma component at all PV coordinates.  To compute the density of N$^{2+}$ and O$^{2+}$ ions in the two plasma components, we adopt the same physical conditions described in the previous section (\S\ref{sec_Hf22_ADF}). 

\begin{figure*}
    \centering
    \includegraphics[width=0.89\linewidth]{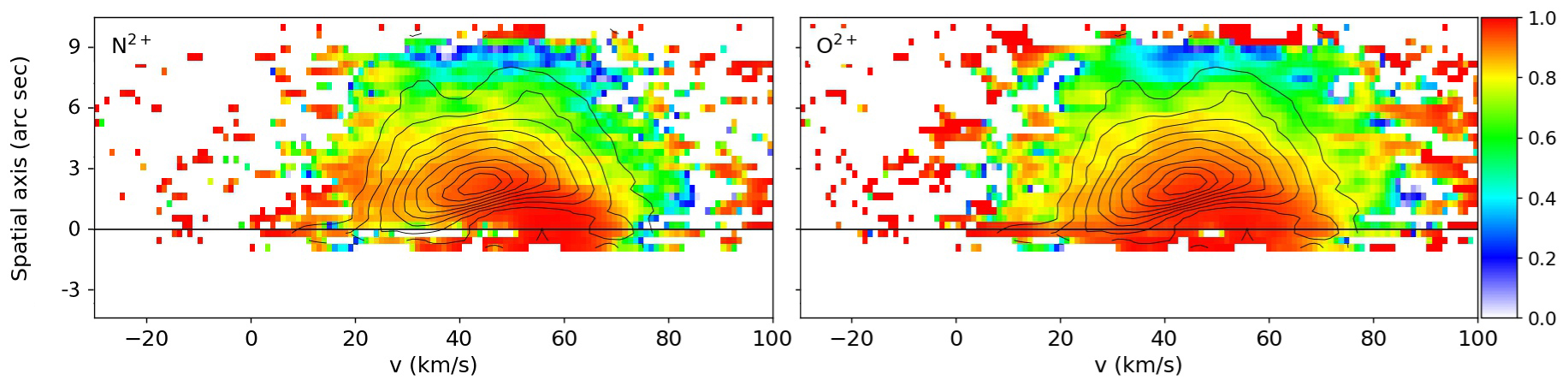}
    \caption{These PV diagrams present the mass fraction of N$^{2+}$ (left) and O$^{2+}$ ions (right) in the additional plasma component for Hf 2-2, calculated under the physical conditions (\te, \nel) = (2,000 K, 5,000 cm$^{-3}$) for the additional plasma component and (\te, \nel) = ([\ion{O}{iii}] \te PV diagram, 1,000 cm$^{-3}$) for the normal nebular plasma. The contour lines correspond to the intensity of the \ion{O}{ii} $\lambda$4649 line.}
    \label{fig:relative-mass-hf22}    
\end{figure*}

Figure \ref{fig:relative-mass-hf22} presents the PV diagrams of the mass fraction of N$^{2+}$ and O$^{2+}$ ions in Hf 2-2 in the additional plasma component. These maps reveal a common spatio-kinematic stratification in the mass fractions.  Along all lines of sight, the mass fraction is minimum at the extreme velocities, though even then accounting for at least $20-50\%$ of the total mass of N$^{2+}$ and O$^{2+}$ ions.  It is noteworthy that the mass fraction in the additional plasma component in Hf 2-2 \emph{never} falls to zero.  In a small volume along the lines of sight near the central star, the additional plasma component contributes nearly \emph{all} of the mass of the N$^{2+}$ and O$^{2+}$ ions.  

\section{M 1-42}

We now turn to M 1-42.  Our analysis in this section follows that already presented for Hf 2-2. We begin by considering the morphology and ionization structure of M 1-42, followed by analyses of the plasma's kinematics, its physical conditions, the oxygen abundances in the two kinematic components, the resulting ADF, and the mass fractions of N$^{2+}$ and O$^{2+}$ ions in the additional plasma component. 

\subsection{Kinematical analysis}

\begin{figure}
    \centering
    \includegraphics[width=\linewidth]{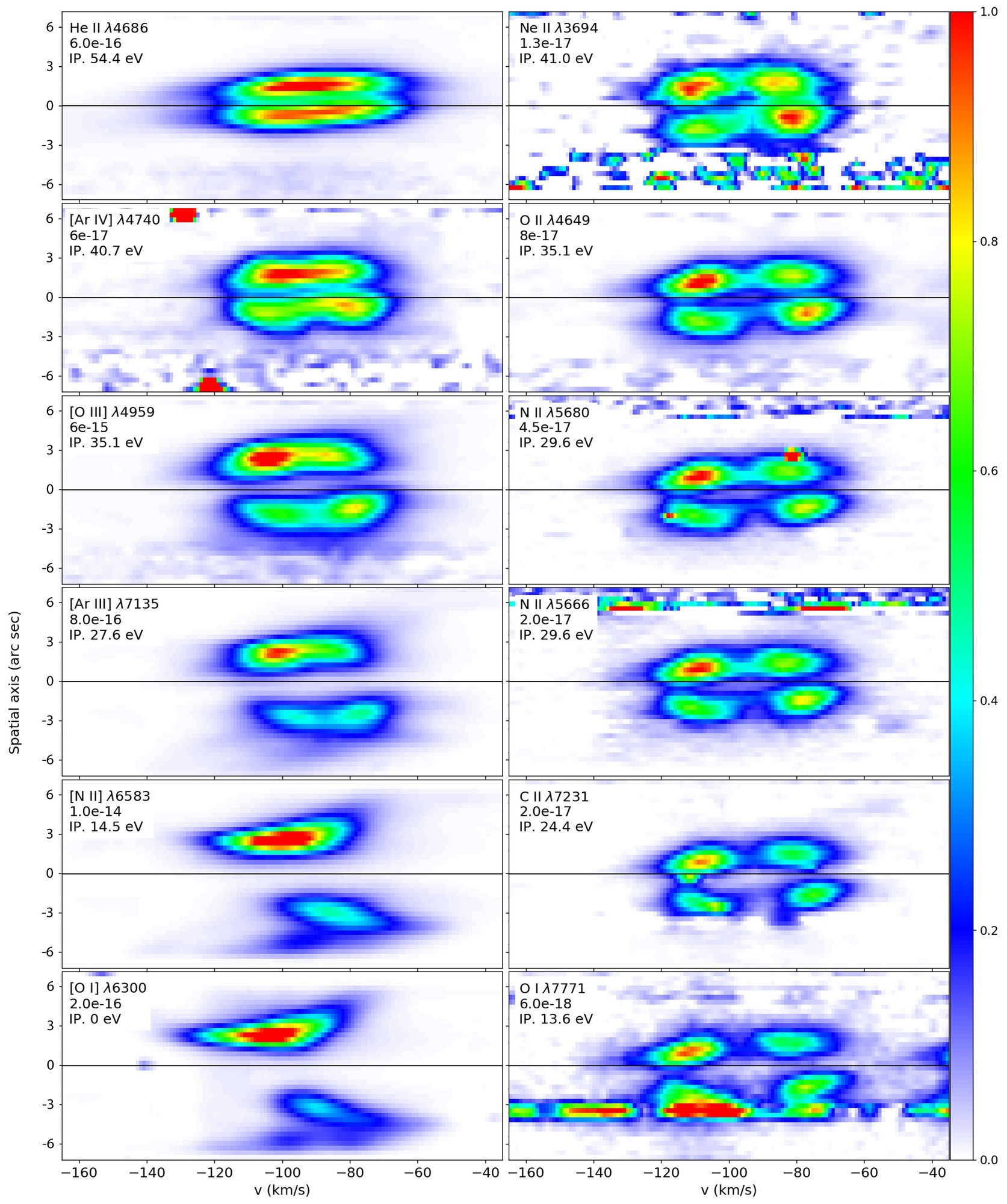}
    \caption{These PV diagrams span the full range of ionization energies of lines observed in M 1-42.  The left column presents the PV diagrams of strong, commonly observed lines while the right column presents those of weak lines of C, N, O, and Ne ions.
    In the PV diagram of \ion{O}{i} \lm7771, there is residual emission from a field star to the SE of M 1-42 (below it in the PV diagram).}
    \label{fig:ion-m142}
\end{figure}

Figure \ref{fig:ion-m142} presents a gallery of PV diagrams of emission lines arranged according to the ionization potential of the ion that produces them. On the left, we present the profiles for \ion{He}{ii} $\lambda$4686 and various forbidden lines.  The right column presents the PV diagrams for permitted lines of \ion{O}{i}, \ion{C}{ii}, \ion{N}{ii}, \ion{O}{ii}, and \ion{Ne}{ii}. The velocity scale is referenced to the systemic velocity of the nebula that we measure, -92 \kms (heliocentric). 

The line profiles in Figure \ref{fig:ion-m142} for M 1-42 differ notably from those of Hf 2-2 (Figure \ref{fig:ion}) in that there is no structure that even approximates a velocity ellipse.  Except in the lines of low ionization ([\ion{N}{ii}] $\lambda$6583, [\ion{O}{i}] $\lambda$6300), there are four emission peaks.  This is not so obvious in the \ion{He}{ii} $\lambda$4686 line, where it also occurs, because it suffers from much greater velocity broadening than the other lines.  On the other hand, all lines of low ionization present a morphology similar to those of [\ion{N}{ii}] $\lambda$6583 and [\ion{O}{i}] $\lambda$6300 that is very different from the four-peak structure of the other lines.  

\begin{figure}
    \centering
    \includegraphics[width=\linewidth]{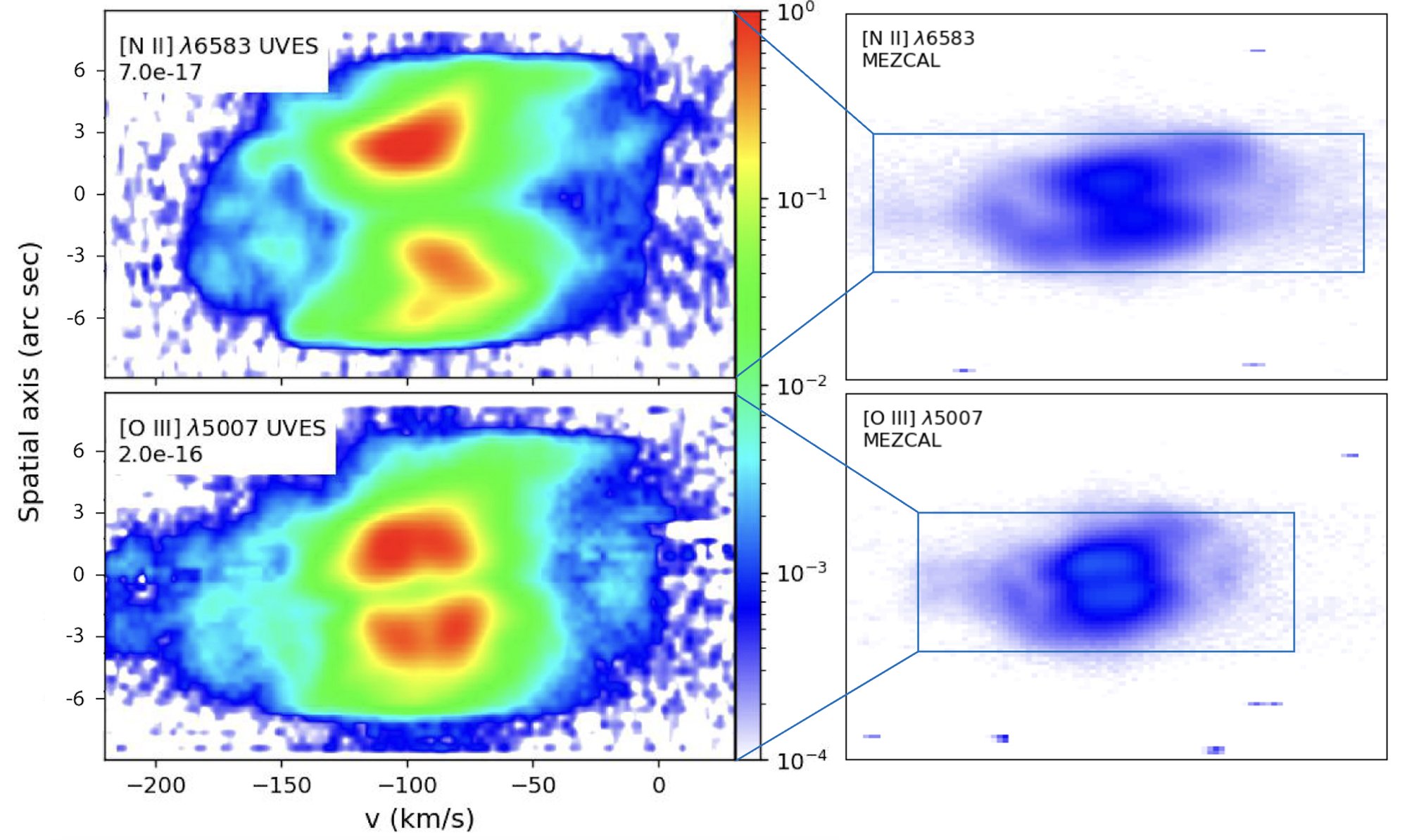}
    \caption{We compare the PV profiles of the [\ion{N}{ii}] $\lambda$6583 (top) and [\ion{O}{iii}] \lm5007 lines (bottom) observed with UVES and the MES-SPM spectrograph. For the MES-SPM observations, the slit has the same position and orientation, but extends over the entire object and so captures emission that falls outside UVES' slit.  The rectangle in the right panels indicates the area included in the UVES PV diagram, which has a greater spatial extent than the UVES slit.  The velocity range of these PV diagrams is considerably greater than those in Figure \ref{fig:ion-m142}.   Here, we use a logarithmic intensity scale to highlight the faintest emission that traces the bipolar lobes of M 1-42.}
    \label{fig:m142_estruc_mezcal}
\end{figure}

The three-dimensional morphology of M 1-42 is better understood from the structure of the faintest emission in the [\ion{O}{iii}] lines and the brightest lines of low ionization.  Figure \ref{fig:m142_estruc_mezcal} compares the PV diagrams of the  [\ion{N}{ii}] $\lambda$ 6583 and [\ion{O}{iii}] $\lambda$5007 lines from our UVES spectroscopy with long slit spectra obtained with the same slit position and orientation using the MES-SPM spectrograph on the 2.1m telescope at the Observatorio Astron\'omico Nacional on the Sierra San Pedro M\'artir.  \citet{lopezetal2012} provide details of the MES-SPM data and their reduction.  Note that the intensity scale of Figure \ref{fig:m142_estruc_mezcal} is logarithmic rather than the linear scale used in previous figures.  The MES-SPM spectra clearly show a bipolar structure in both lines, with the (faint) lobes more or less aligned along the line of sight.  This is less obvious in the UVES spectra, because its slit does not extend over the full spatial extent of M 1-42.  

\begin{figure*}
    \centering
    \includegraphics[width=\linewidth]{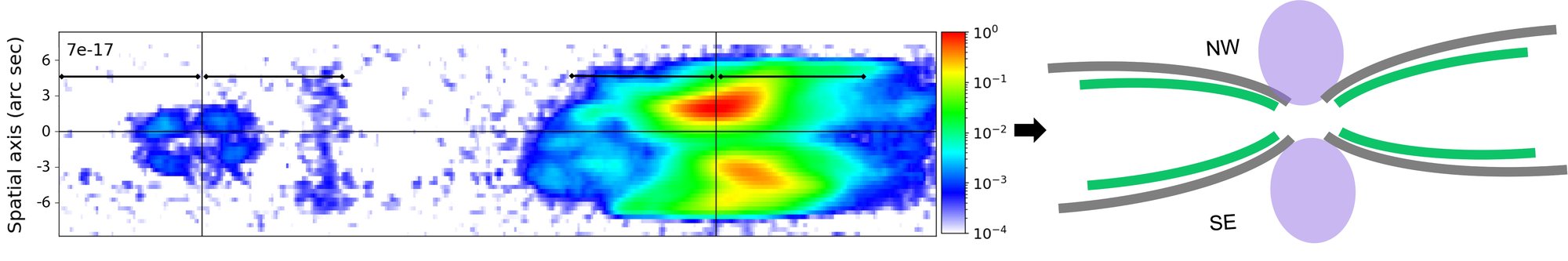}
    \caption{Left:  We present the region of the \ion{C}{ii} $\lambda$6578 and [\ion{N}{ii}] $\lambda$6583 lines (UVES spectrum) using a logarithmic intensity scale to display the faintest structures.  The vertical lines indicate the rest wavelengths for the two lines at the systemic velocity of M 1-42.  The horizontal line is the position of the central star.  The ``error bars" centred at the rest wavelengths indicate the 130\,km/s interval used for all other PV diagrams (except Figure \ref{fig:m142_estruc_mezcal}).  The velocity range of [\ion{N}{ii}] $\lambda$6583 greatly exceeds that of \ion{C}{ii} $\lambda$6478.   Right:  We present a schematic diagram of the structure of M 1-42. The arrow indicates the line of sight.  The purple ovals represent the torus surrounding the central star and produce the  emission seen in the lines of low ionization (Figure \ref{fig:ion-m142}, [\ion{N}{ii}] $\lambda$6583, [\ion{O}{i}] $\lambda$6300).  The gray lines represent the lobes that emit the normal nebular plasma, best observed in lines of intermediate and high ionization (Figure \ref{fig:ion-m142}, upper four panels, left column).  The green lines represent the additional plasma component that is observed in lines of \ion{O}{i}, \ion{C}{ii}, \ion{N}{ii}, \ion{O}{ii}, and \ion{Ne}{ii} (Figure \ref{fig:ion-m142}, right column).   }
    \label{fig:lobulos}
\end{figure*}

Figure \ref{fig:lobulos} presents the spectral region of the \ion{C}{ii} $\lambda$6578 and [\ion{N}{ii}] $\lambda$6583 lines using a logarithmic intensity scale to show the faintest emission.  With this intensity scale, it is clear that the velocity extent of the \ion{C}{ii} $\lambda$6578 line is much less than that of [\ion{N}{ii}] $\lambda$6583.  The \ion{C}{ii} $\lambda$6578 emission is also confined to a spatial region closer to the line of sight to the central star, a trait it shares with the lines of \ion{O}{i}, \ion{N}{ii}, and \ion{O}{ii}\ (Figure \ref{fig:ion-m142}).  Therefore, these lines easily fit, both spatially and in velocity, within the lobes defined by the [\ion{N}{ii}] $\lambda$6583 line. Thus, the three-dimensional structure of M 1-42 is likely that of a thick toroid or disc from which the ionization due to the central star is inflating a pair of expanding (broken?) lobes.  We attempt to illustrate this structure in the right panel of Figure \ref{fig:lobulos}.  Given the similar structure of the PV diagrams of the lines of lowest ionization, that are distinctly different from those of higher ionization (Figure \ref{fig:ion-m142}), it is likely that the toroid in M 1-42 is optically thick to ionizing radiation while the lobes are optically thin.  

For a bipolar structure seen pole-on, we would normally study its expansion using the velocities of the end caps of the two lobes.  However, these are not seen in the case of M 1-42, except in low ionization lines and [\ion{O}{iii}]  $\lambda\lambda$4959,5007. Instead, we measured the velocity separation of emission peaks to the NW and SE of the central star, which presumably reflects the plasma motion with respect to the toroid.  The drawback to this approach of characterizing the kinematics is that it precludes using information from the lines of low ionization ions.  Since there is no velocity ellipse in the PV diagramas for M 1-42, we measured the velocity splitting at the spatial position of the peak emission in the NW and SE knots, rather than at some fixed spatial position. Table \ref{tab:m1vel} presents the resulting velocity differences.  

\begin{figure*}
    \centering
    \includegraphics[width=0.89\linewidth]{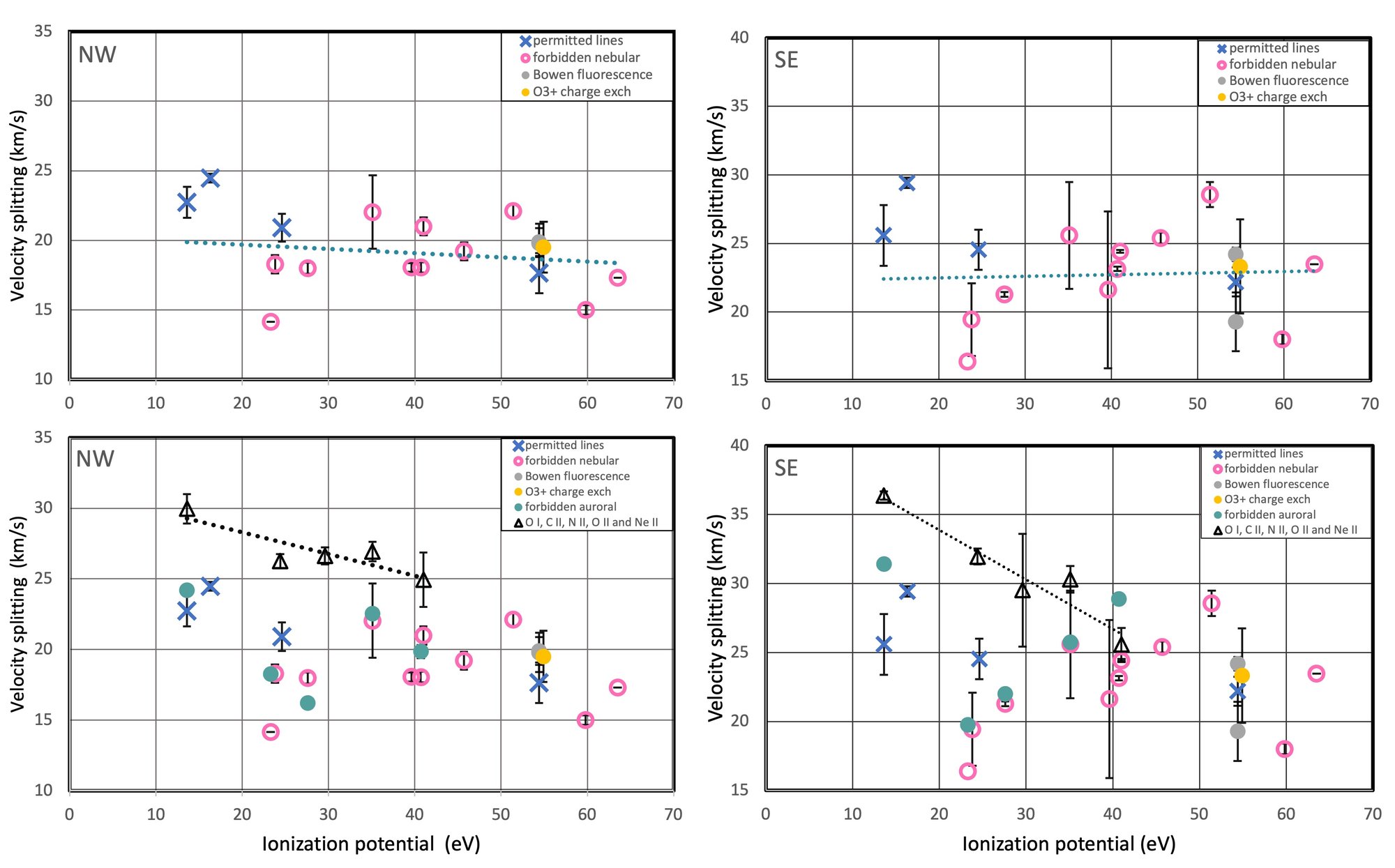}
    \caption{We present the Wilson diagrams for the northwest (left) and southeastern regions (right) in M 1-42. The panels in the top row include the forbidden nebular transitions and lines due to fluorescence, recombination, and charge exchange. The panels in the bottom row add the forbidden auroral lines and the lines of \ion{O}{i}, \ion{C}{ii}, \ion{N}{ii},\ion{O}{ii} and \ion{Ne}{ii}. The dotted lines are linear fits to the lines emitted by the normal nebular plasma (top row) and the additional plasma component (bottom row).}
    \label{fig:m142-kinematical}
\end{figure*}

We present the results for the NW and SE emission peaks separately 
in Figure \ref{fig:m142-kinematical}. In the top row, we present the velocity splitting of the nebular forbidden lines and lines due to recombination, fluorescence, and charge exchange.  For both the NW and SE emission peaks, the velocity splitting shows no systematic trend with respect to the ionization potential.  The linear fit to all of the lines is compatible with a slope of zero in both cases.  However, as Figure \ref{fig:ion-m142} indicates (left column), the spatial separation between the NW and SE emission peaks increases systematically as the ionization potential decreases, so what we are likely observing is a radial ionization structure in the surface of the central toroid.

On the other hand, the lines of \ion{O}{i}, \ion{C}{ii}, \ion{N}{ii}, and \ion{O}{ii} present a different behaviour.  For both the NW and SE emission peaks, there is a Wilson-like correlation between the velocity splitting and the ionization potential, with greater velocity splitting at lower ionization levels (Figure \ref{fig:m142-kinematical}, bottom row).  A linear fit to these data indicates a slope that is statistically different from zero.  As well, for both the NW and SE emission peaks, the velocity splitting of these permitted lines is greater than those for the lines shown in the top row.  The emission from these lines is also observed closer to the line of sight towards the central star (Figure \ref{fig:ion-m142}, right column). 

We attempt to relate the kinematics of M 1-42 to its possible structure in the right panel of Figure \ref{fig:lobulos}.  Most of the emission from [\ion{N}{ii}] $\lambda$6583 and other low ionization lines arises within the volume of the central toroid/disc, shown in purple, but a minority from the lobes, shown with the gray line.  It is possible that most of the toroid/disc is not ionized.  Note that the lobes may not be closed or complete, since they do not present a sharp border in velocity in Figure \ref{fig:m142_estruc_mezcal}.  They may simply be matter flowing away from the toroid.  For that reason, we draw the lobes as open.  The emission from lines of higher ionization could arise from where the lobes connect to the toroid/disc, as seen by the central star (the gray line).  Finally, the emission from the lines of \ion{O}{i}, \ion{C}{ii}, \ion{N}{ii}, and \ion{O}{ii} arise interior to the lobes, shown schematically with the green lines.  

\begin{table}
\centering
\caption{Ionization potential and velocity splitting of ions observed for M 1-42 (separated by regions)}
\resizebox{0.48\textwidth}{!}{
\begin{tabular}{ccccl}
\toprule   
  Parent       & E$_{ion}$    & \multicolumn{2}{c}{Velocity Splitting} & \multicolumn{1}{c}{Lines (\AA)}                                               \\  
                     &(eV)&  \multicolumn{2}{c}{(km~s$^{-1}$)} & \\
                     && Northwest & Southeast&
\\
\midrule
\begin{tabular}[c]{@{}l@{}}H$^{+}$ \\ \\\end{tabular}      & \begin{tabular}[c]{@{}l@{}} 13.6 \\ \\ \end{tabular}       &  \begin{tabular}[c]{@{}l@{}} 22.7 $\pm$ 1.1          \\ \\ \end{tabular}  & \begin{tabular}[c]{@{}l@{}} 25.6 $\pm$ 2.2       \\ \\ \end{tabular}  & \begin{tabular}[c]{@{}l@{}}\ion{H}{i} $\lambda \lambda$3770, 3797, 3835, 3970, 9229, \\ 9545, 10049   \end{tabular}                            \\
 \begin{tabular}[c]{@{}l@{}} He$^{+}$ \\ \\ \\  \end{tabular}  & \begin{tabular}[c]{@{}l@{}} 24.6  \\ \\ \\  \end{tabular}      & \begin{tabular}[c]{@{}l@{}} 21.0 $\pm$ 1.0    \\ \\ \\  \end{tabular}      & \begin{tabular}[c]{@{}l@{}} 24.5 $\pm$ 1.5  \\ \\ \\  \end{tabular}         & \begin{tabular}[c]{@{}l@{}} \ion{He}{i}$\lambda \lambda$3187, 3614, 3065, 4026, 4387, \\ 4471, 4713, 4922, 5015, 5875, 6678, \\ 7281 \end{tabular} \\
\begin{tabular}[c]{@{}l@{}} He$^{2+}$ \\ \\ \end{tabular}   & \begin{tabular}[c]{@{}l@{}} 54.4  \\ \\ \end{tabular}      & \begin{tabular}[c]{@{}l@{}} 17.6 $\pm$ 1.4      \\ \\ \end{tabular}    & \begin{tabular}[c]{@{}l@{}} 22.2 $\pm$ 1.1     \\ \\ \end{tabular}    & \begin{tabular}[c]{@{}l@{}} \ion{He}{ii} $\lambda \lambda$4541, 4686, 4859, 5411, 6560, \\  10123  \end{tabular}                                 \\
C$^{2+}$     & 24.4        & 26.3 $\pm$ 0.5          & 32.0 $\pm$ 0.6         & \ion{C}{ii} $\lambda \lambda$5342, 6461, 6578, 7231                                                 \\
\begin{tabular}[c]{@{}l@{}} N$^{2+}$ \\ \\ \end{tabular}     & \begin{tabular}[c]{@{}l@{}} 29.6   \\ \\ \end{tabular}     & \begin{tabular}[c]{@{}l@{}} 26.6 $\pm$ 0.6    \\ \\ \end{tabular}      & \begin{tabular}[c]{@{}l@{}} 29.5 $\pm$ 4.1    \\ \\ \end{tabular}     & \begin{tabular}[c]{@{}l@{}} \ion{N}{ii} $\lambda \lambda$4035, 4041, 5666, 5676, 5680,\\  5710  \end{tabular}                          \\
N$^{2+}$     & 54.4        & 19.9 $\pm$ 1.0          & 24.2 $\pm$ 0.5         & \ion{N}{iii} $\lambda \lambda$4097, 4634, 4641 (Bowen fl.)                                                \\
O$^{+}$      & 13.6        & 24.2                    & 31.4                   & {[}\ion{O}{ii}{]} $\lambda \lambda$7319 (auroral)                                                             \\
O$^{+}$      & 13.6        & 30.0 $\pm$ 1.0          & 36.4 $\pm$ 0.3         & \ion{O}{i} $\lambda \lambda$7771, 9265                                                              \\
O$^{2+}$     & 35.1        & 22.0 $\pm$ 3            & 25.6 $\pm$ 3.9         & {[}\ion{O}{iii}{]} $\lambda$4959 (nebular)                                                                    \\
O$^{2+}$     & 35.1        & 22.5                    & 25.7                   & {[}\ion{O}{iii}{]} $\lambda$4363 (auroral)                                                                    \\
\begin{tabular}[c]{@{}l@{}} O$^{2+}$    \\ \\ \end{tabular} & \begin{tabular}[c]{@{}l@{}} 35.1   \\ \\ \end{tabular}     & \begin{tabular}[c]{@{}l@{}} 26.9 $\pm$ 0.7    \\ \\ \end{tabular}      & \begin{tabular}[c]{@{}l@{}} 30.3 $\pm$ 1.0     \\ \\ \end{tabular}     & \begin{tabular}[c]{@{}l@{}}\ion{O}{ii} $\lambda \lambda$4072, 4084, 4089,  4349, 4366,\\ 4639, 4649, 4651, 4662, 4674, 4676    \end{tabular}  \\
O$^{2+}$     & 54.4        & 19.8 $\pm$ 1.4          & 19.3 $\pm$ 2.1         &\ion{O}{iii} $\lambda \lambda$3133, 3754, 3759 (Bowen fl.)                                                \\
O$^{3+}$     & 54.9        & 19.5 $\pm$ 1.8          & 23.3 $\pm$ 3.4         &\ion{O}{iii} $\lambda \lambda$3757, 3774 (charge exchange)                                                      \\
Ne$^{2+}$    & 41.0        & 21.0 $\pm$ 0.6          & 24.4 $\pm$ 0.1         & {[}\ion{Ne}{iii}{]} $\lambda \lambda$3869, 3967 (nebular)                                                     \\
Ne$^{2+}$    & 41          & 25.0 $\pm$ 1.9          & 25.6 $\pm$ 1.2         & \ion{Ne}{ii}  $\lambda$3694                                                                         \\ 
Ne$^{3+}$    & 63.5        & 17.3                    & 23.5                   & {[}\ion{Ne}{iv}{]} $\lambda \lambda$4724 (nebular)                                                            \\
Si$^{2+}$    & 16.3        & 24.5 $\pm$ 0.3          & 29.4 $\pm$ 0.4         & \ion{Si}{ii} $\lambda \lambda$5041, 6347, 6371                                                      \\
S$^{2+}$     & 23.3        & 14.1                    & 16.4                   & {[}\ion{S}{iii}{]} $\lambda$9531 (nebular)                                                                    \\
S$^{2+}$     & 23.3        & 18.3 $\pm$ 0.04         & 19.8 $\pm$ 0.2         & {[}\ion{S}{iii}{]} $\lambda$6312 (auroral)                                                                    \\
Cl$^{2+}$    & 23.8        & 18.3 $\pm$ 0.7          & 19.4 $\pm$ 2.6         & {[}\ion{Cl}{iii}{]} $\lambda \lambda$5517, 5537 (nebular)                                                     \\
Cl$^{3+}$    & 39.6        & 18.1 $\pm$ 0.3          & 21.6 $\pm$ 5.7         & {[}\ion{Cl}{iv}{]} $\lambda \lambda$7530, 8045 (nebular)                                                      \\
Ar$^{2+}$    & 27.6        & 18.0 $\pm$ 0.4          & 21.3 $\pm$ 0.2         & {[}\ion{Ar}{iii}{]} $\lambda \lambda$7135, 7751 (nebular)                                                     \\
Ar$^{2+}$    & 27.6        & 16.2                    & 22.0                   & {[}\ion{Ar}{iii}{]} $\lambda$5191 (auroral)                                                                   \\
Ar$^{3+}$    & 40.7        & 18.0 $\pm$ 0.4          & 23.1 $\pm$ 0.2         & {[}\ion{Ar}{iv}{]} $\lambda \lambda$4711, 4740 (nebular)                                                      \\
Ar$^{3+}$    & 40.7        & 19.9 $\pm$ 0.5          & 28.9 $\pm$ 0.01        & {[}\ion{Ar}{iv}{]} $\lambda$7262 (nebular)                                                                    \\
Ar$^{4+}$    & 59.8        & 15.0 $\pm$ 0.3          & 18.01 $\pm$ 0.4        & {[}\ion{Ar}{v}{]} $\lambda \lambda$6435, 7006 (nebular)                                                       \\
K$^{3+}$     & 45.7        & 19.2 $\pm$ 0.6          & 25.4 $\pm$ 0.4         & {[}\ion{K}{iv}{]} $\lambda \lambda$6101, 6795 (nebular)                                                       \\
Mn$^{4+}$    & 51.4        & 22.0 $\pm$ 0.5          & 27.7 $\pm$ 0.9         & {[}\ion{Mn}{v}{]} $\lambda \lambda$5701, 5861, 6083, 6393  \\
\bottomrule   
\end{tabular}
}
\label{tab:m1vel}
\end{table}

\subsection{Physical Conditions}
\subsubsection{Forbidden lines}

We compute the electron densities and temperatures in M 1-42 as for Hf 2-2 (\S\ref{sub:phycon}).  We begin with the forbidden lines and then consider the lines of \ion{N}{ii} and \ion{O}{ii} in the next subsection.  

\begin{figure*}
    \centering
    \includegraphics[width=0.89\linewidth]{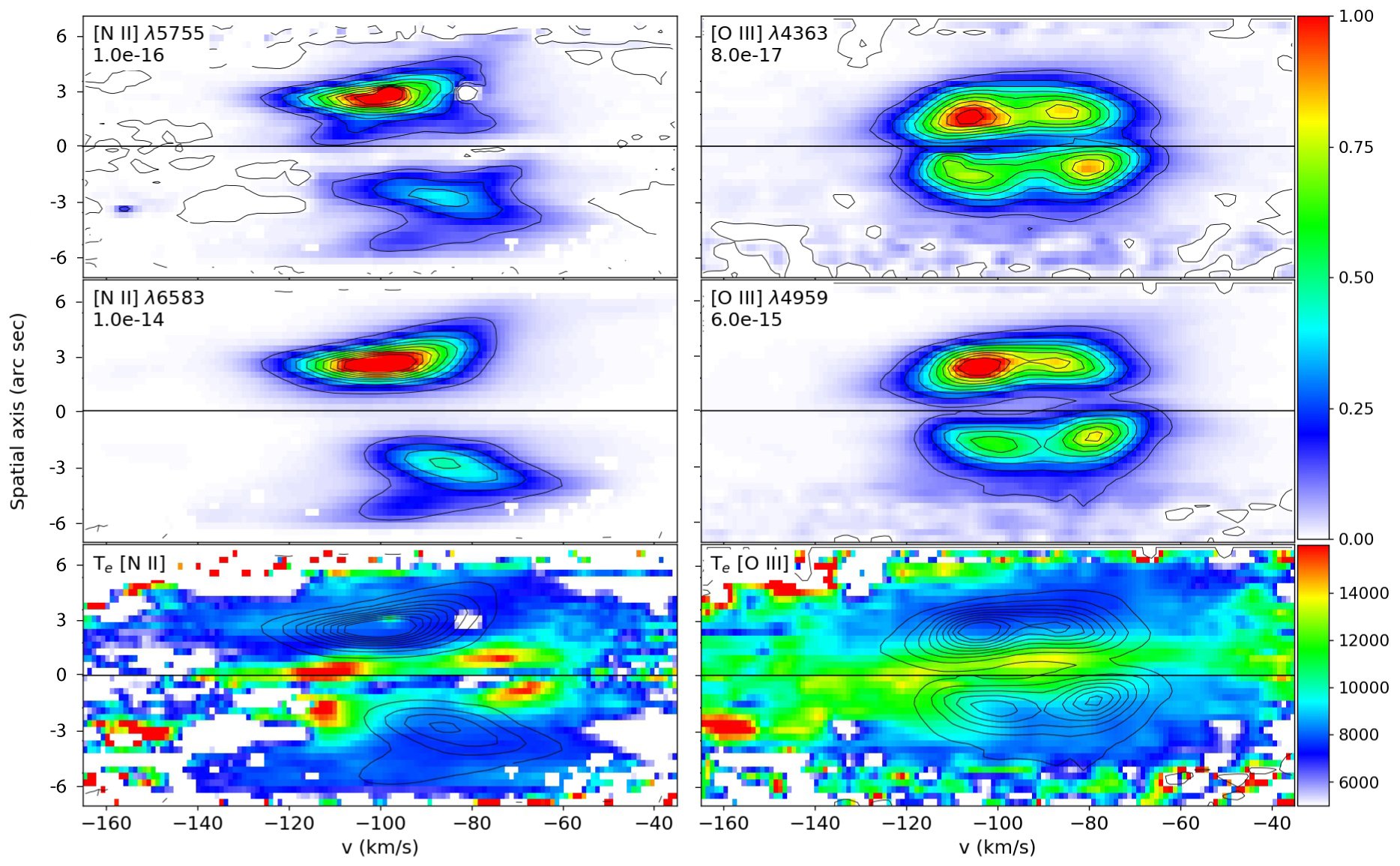}
    \caption{In the top two rows, we present the PV diagrams for M 1-42 of the [\ion{N}{ii}] $\lambda\lambda$5755,6583 (left) and [\ion{O}{iii}] $\lambda\lambda$4363,4959 lines (right) and of the electron temperatures derived from them (bottom row), assuming $n_e = 1,000\,\mathrm{cm}^{-3}$. The contours in the bottom row correspond to the intensity of [\ion{N}{ii}] $\lambda$6583 (left) and [\ion{O}{iii}] $\lambda$4959 lines (right).  For the [\ion{N}{ii}] lines, we attempted to subtract the contribution due to recombination, which was again only partly successful.}
    \label{fig:tem[oiii-nii]-m142}
\end{figure*}

Figure \ref{fig:tem[oiii-nii]-m142} (right) shows the PV diagrams for [\ion{O}{iii}] $\lambda\lambda$4363,4959. These lines present the four knots on both sides of the central torus. We compute the electron temperature shown in the bottom-right panel of Figure \ref{fig:tem[oiii-nii]-m142}, adopting an electron density of 1,000 cm$^{-3}$. The temperature profile again shows a marked gradient in the spatial direction, with a temperature of around 13,000 K near the star and approximately 7,000 K farther away. No significant temperature variations are observed along the line of sight (i.e., at different velocities at the same distance from the star).

To assess whether the [\ion{O}{iii}] lines are contaminated by recombination and charge exchange processes, we present the PV diagrams of the recombination line \ion{O}{iii} $\lambda$3265 and the charge exchange line \ion{O}{iii} $\lambda$5592 in Figure \ref{fig:reclines-m142}. The intensities of these lines are about 100 times weaker than the peak intensity of [\ion{O}{iii}] $\lambda$4363. Thus, contamination of the [\ion{O}{iii}] $\lambda$4363 from these processes is minor, at most.

We also computed the electron temperature in M 1-42 using the PV diagrams of [\ion{N}{ii}] $\lambda\lambda$5755,6583 (Figure \ref{fig:tem[oiii-nii]-m142}, left column), after correcting them for the contribution due to recombination as described for Hf 2-2 (\S\ref{sub:forl}). See Figures \ref{fig:m142-nii-oii-decom} and \ref{fig:m142-decon} in the Appendix for the intermediate steps.  The line profiles for these lines are typical of those of lower ionization lines (Figure \ref{fig:ion-m142}), lacking the four-knot morphology of the higher ionization lines.  Note that, even after correcting for the contribution due to recombination, the morphology of the PV diagram of [\ion{N}{ii}] $\lambda$5755 differs from that of $\lambda$6583, as there is still residual structure resembling the four knots observed in the PV diagrams of permitted lines in Figure \ref{fig:ion-m142}. (See also Figure \ref{fig:m142-decon}.) Even so, the resulting temperatures in the PV map are not so different from the temperature map based upon the [\ion{O}{iii}] lines.  In the outer regions of the object, where there is no contribution from recombination, the temperature is around 8,000 K.  Closer to the central star, the electron temperature rises to values similar to those found in [\ion{O}{iii}], but the structure is slightly different.  Hence, we find rather good agreement in the electron temperature maps from [\ion{N}{ii}] and [\ion{O}{iii}]. 

\begin{figure*}
    \centering
    \includegraphics[width=0.89\linewidth]{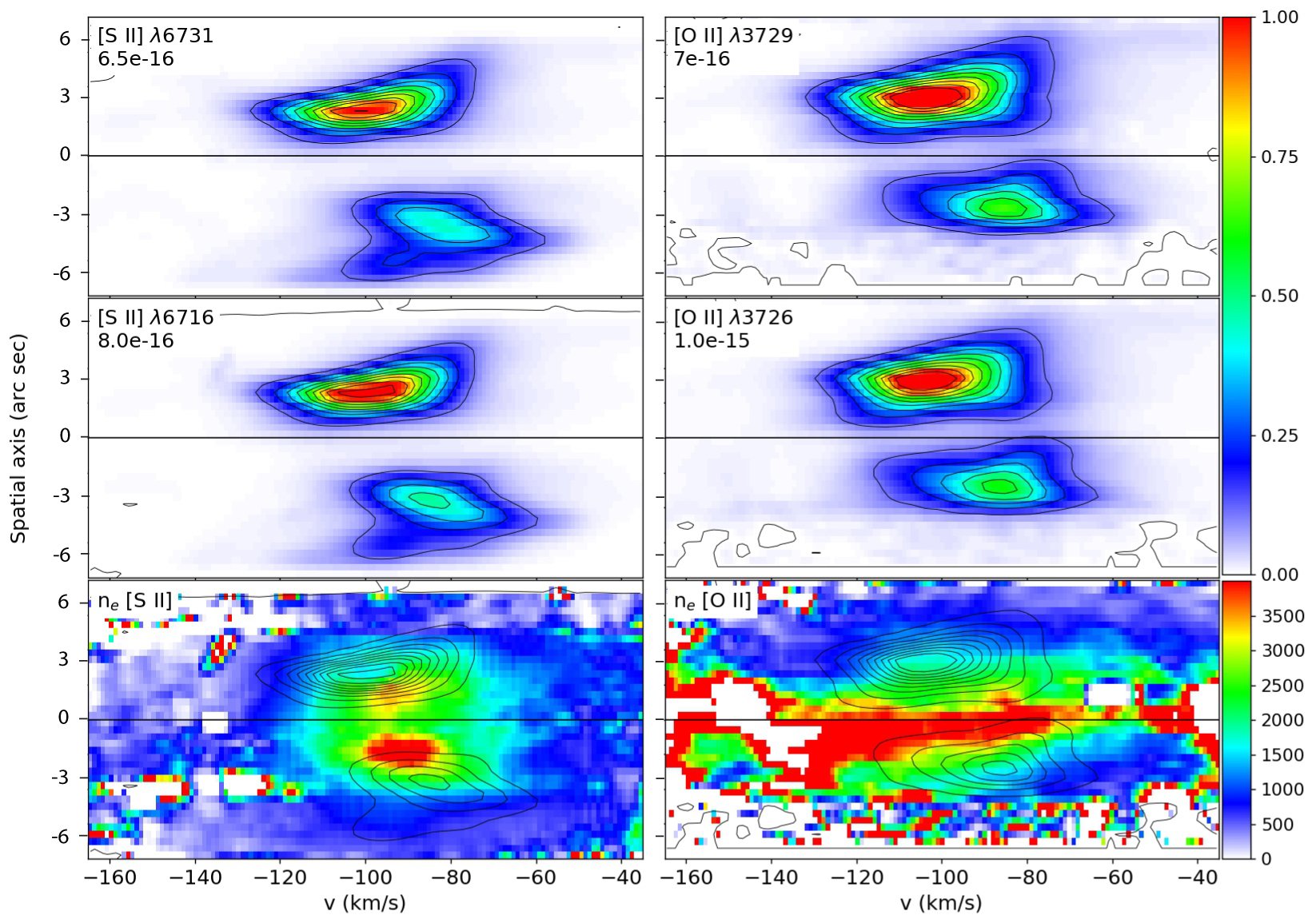}
    \caption{The top two rows present the PV diagrams for M 1-42 of the [\ion{S}{ii}] $\lambda\lambda$6716,6731 (left) and [\ion{O}{ii}] $\lambda\lambda$3726,3729 lines (right) and the electron density derived from them (bottom row), assuming $T_e = 10,000\,\mathrm K$. The contours in the bottom row correspond to the intensity of [\ion{S}{ii}] $\lambda$6731 (left) and [\ion{O}{ii}] $\lambda$3726 lines (right).  The very different density structures likely reflect the contribution of recombination to the [\ion{O}{ii}] lines.}
    \label{fig:den[sii-oii]-m142}
\end{figure*} 

In Figure \ref{fig:den[sii-oii]-m142} (left column), we present the PV diagrams of the [\ion{S}{ii}] $\lambda \lambda$6716,6731 lines for M 1-42 as well as the electron density derived from them, assuming a temperature of 10,000 K. There are two density peaks near the systemic velocity where the electron density exceeds 3,000 cm$^{-3}$.  These density peaks do not coincide with the peaks in the intensities of the [\ion{S}{ii}] lines used for the diagnostic.  Presumably, these peaks correspond to the central torus.  At velocities that differ substantially from the systemic velocity and along lines of sight farther from the central star, we find much lower densities, of order 500 cm$^{-3}$. 

We also compute the electron density using the [\ion{O}{ii}] $\lambda\lambda$3729,3726 lines (Figure \ref{fig:den[sii-oii]-m142}, right column).  The morphology of these PV diagrams differs slightly from those of the [\ion{S}{ii}] $\lambda\lambda$6716,6731 or [\ion{N}{ii}] $\lambda$6583 lines in that there is much more emission on lines of sight close to the central star.  The density map derived from the [\ion{O}{ii}] $\lambda\lambda$3729,3726 lines differs significantly from that for the [\ion{S}{ii}] $\lambda\lambda$6716,6731 lines in that there is a region of high density, exceeding 4,000 cm$^{-3}$, covering a wide velocity range along lines of sight very close to the central star.  Along lines of sight farther from the central star, the electron density is similar to that found from the [\ion{S}{ii}] $\lambda\lambda$6716,6731 lines.  

The difference in the electron density indicated by the [\ion{S}{ii}] and [\ion{O}{ii}] lines in M 1-42 may be understood if the [\ion{O}{ii}] lines are affected by a contribution due to recombination.  In that case, as in Hf 2-2, the [\ion{O}{ii}] lines may reflect the density in the plasma from which the permitted \ion{O}{ii} lines are emitted.  The permitted lines have higher velocities compared to the systemic velocity and so could explain the larger range in velocity observed in the density map derived from the [\ion{O}{ii}] lines.

\subsubsection{Permitted lines} \label{sec_m142_physcond_permitted}

\begin{figure*}
    \centering
    \includegraphics[width=0.89\linewidth]{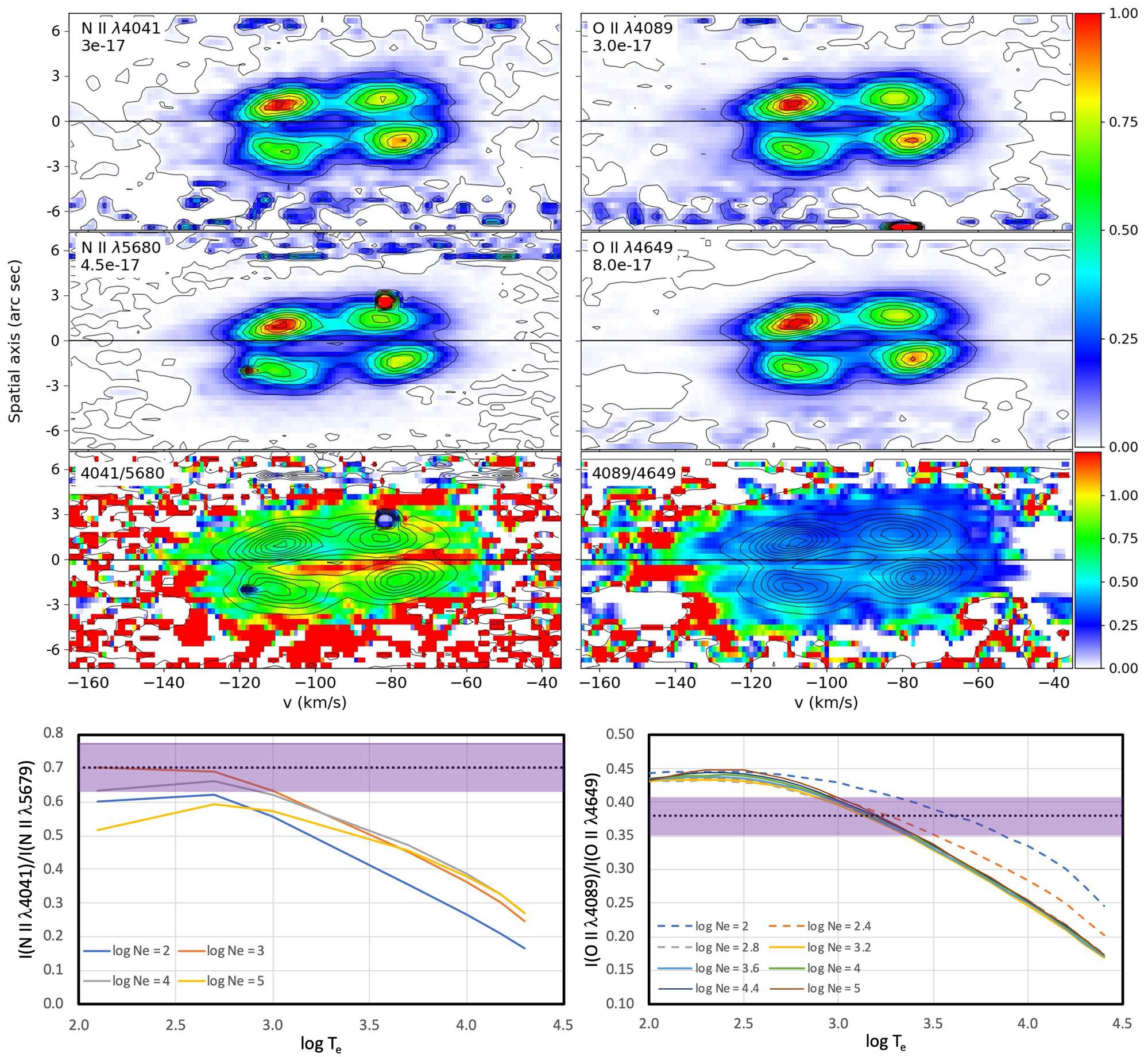}
    \caption{The top two rows present the PV diagrams of the \ion{N}{ii} $\lambda\lambda$4041,5680 (left) and \ion{O}{ii} $\lambda\lambda$4089,4649 lines (right) that we use to determine \te in M 1-42.  The third row displays the PV diagrams of the line ratios derived from the previous two rows, with the isocontours of the intensity of the \ion{N}{ii} $\lambda$5680 (right) and \ion{O}{ii} $\lambda$4649 lines (right) overlaid, respectively. The bottom row presents the line ratio as a function of \te for different \nel values (Table \ref{tab:atomic-data}). The mean value (dotted line) and standard deviation (purple shading), estimated from the PV diagrams in the third row, considering the area interior to the 40\% contour, but excluding the two cosmic rays that contaminate the \ion{N}{ii} $\lambda$5680 line.}
    \label{fig:temrec_nii_oii-m142}
\end{figure*}

In Figure \ref{fig:temrec_nii_oii-m142}, we present the electron temperature we calculate for M 1-42 using the permitted lines of \ion{N}{ii} and \ion{O}{ii}. In the left column, we present our analysis based upon the \ion{N}{ii} lines.  The third panel shows the line ratio, which is relatively flat without significant gradients.  Considering only those pixels with intensities above 40$\%$ of the maximum intensity in the PV diagram of $\lambda$5680, we obtained a mean \ion{N}{ii} $\lambda\lambda$4041/5680 ratio of $0.70 \pm 0.07$.  The diagnostic diagram in the bottom row of Figure \ref{fig:temrec_nii_oii-m142} indicates that this intensity ratio implies a temperature below approximately 1,000\,K.

We find a similar result using the \ion{O}{ii} $\lambda\lambda$4089,4649 lines.  Figure \ref{fig:temrec_nii_oii-m142} (right column) presents the PV diagrams and the line ratio.  As observed for the \ion{N}{ii} recombination lines, the line ratio (third panel) is uniform.  Its average value is $0.38 \pm 0.03$, based upon all pixels exceeding 40\% of the maximum intensity in the PV diagram of $\lambda$4649. The diagnostic diagram for these lines (bottom panel, right) indicates that this value implies an electron temperature in the range of $600-3,000$\,K, unless the density is very low. 

\begin{figure*}
    \centering
    \includegraphics[width=0.89\linewidth]{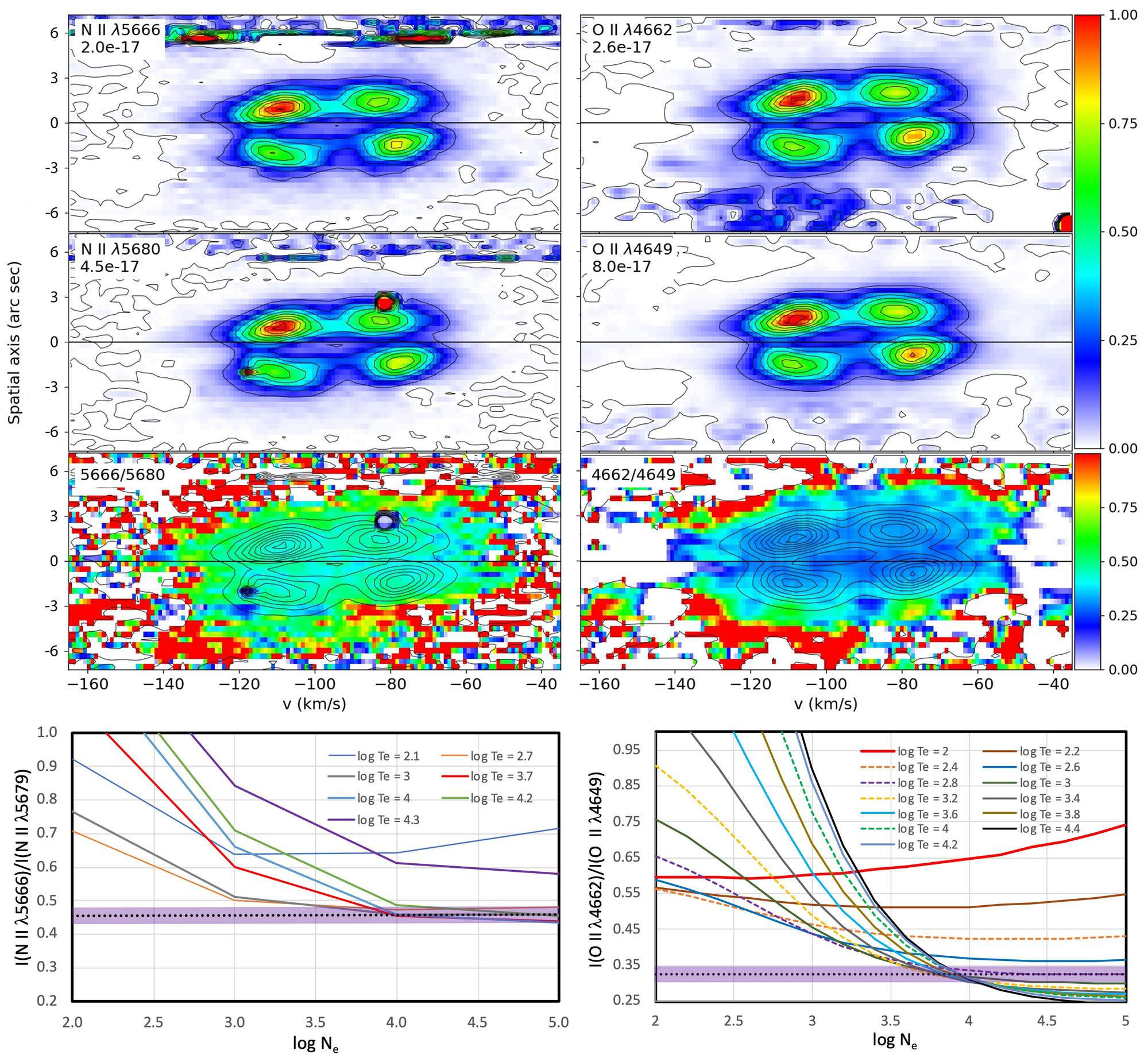}
    \caption{We present the PV diagrams of the \ion{N}{ii} $\lambda\lambda$5666,5680 (left) and \ion{O}{ii} $\lambda\lambda$4649,4662 lines (right) used to determine \nel in M 1-42 in the top two rows.
    The third row displays the line ratios with the isocontours of the intensity of the \ion{N}{ii} $\lambda$5680 and \ion{O}{ii} $\lambda$4649 lines overlaid. The bottom row shows the line ratio as a function of \te for different \nel values (Table \ref{tab:atomic-data}).  We indicate the mean value (dotted line) and standard deviation (purple shading), estimated from the line ratios in the third row, using the area within the 40\% contour, but excluding the two cosmic rays that contaminate the \ion{N}{ii} $\lambda$5680 line.  }
    \label{fig:ne_m142_relines}
\end{figure*}

As regards the electron density in M 1-42, the analysis based upon the \ion{N}{ii} and \ion{O}{ii} lines is found in Figure \ref{fig:ne_m142_relines}.  The left column presents the PV diagrams and the line ratio for the \ion{N}{ii} $\lambda\lambda$5666/5680 lines. Again, using only those pixels that exceed 40\% of the maximum value in the \ion{N}{ii} $\lambda$5680 line, we find an average value of $0.46\pm 0.02$  which implies a density exceeding 3,000\,cm$^{-3}$. The same analysis based upon the \ion{O}{ii} $\lambda\lambda$4649,4662 lines ($0.32\pm 0.02$) implies an electron density in excess of $4,000$\,cm$^{-3}$ and an electron temperature in the $400-1,000$\,K range.  

Table \ref{summary}  summarizes the \te and \nel values derived from the recombination lines.  Considering the joint constraints from the \ion{N}{ii} and \ion{O}{ii} lines, for the additional plasma component in M 1-42 we find $n_e > 4,000$\,cm$^{-3}$ and $400< T_e < 1,000$\,K.  The electron density is compatible with the value found for the [\ion{O}{ii}] lines.  As was found for Hf 2-2, the additional plasma component in M 1-42 is colder and denser than the normal nebular plasma.  

\subsection{Ionic Abundances and ADF}\label{sec_m142ADF}

\begin{figure}
    \centering
    \includegraphics[width=\linewidth]{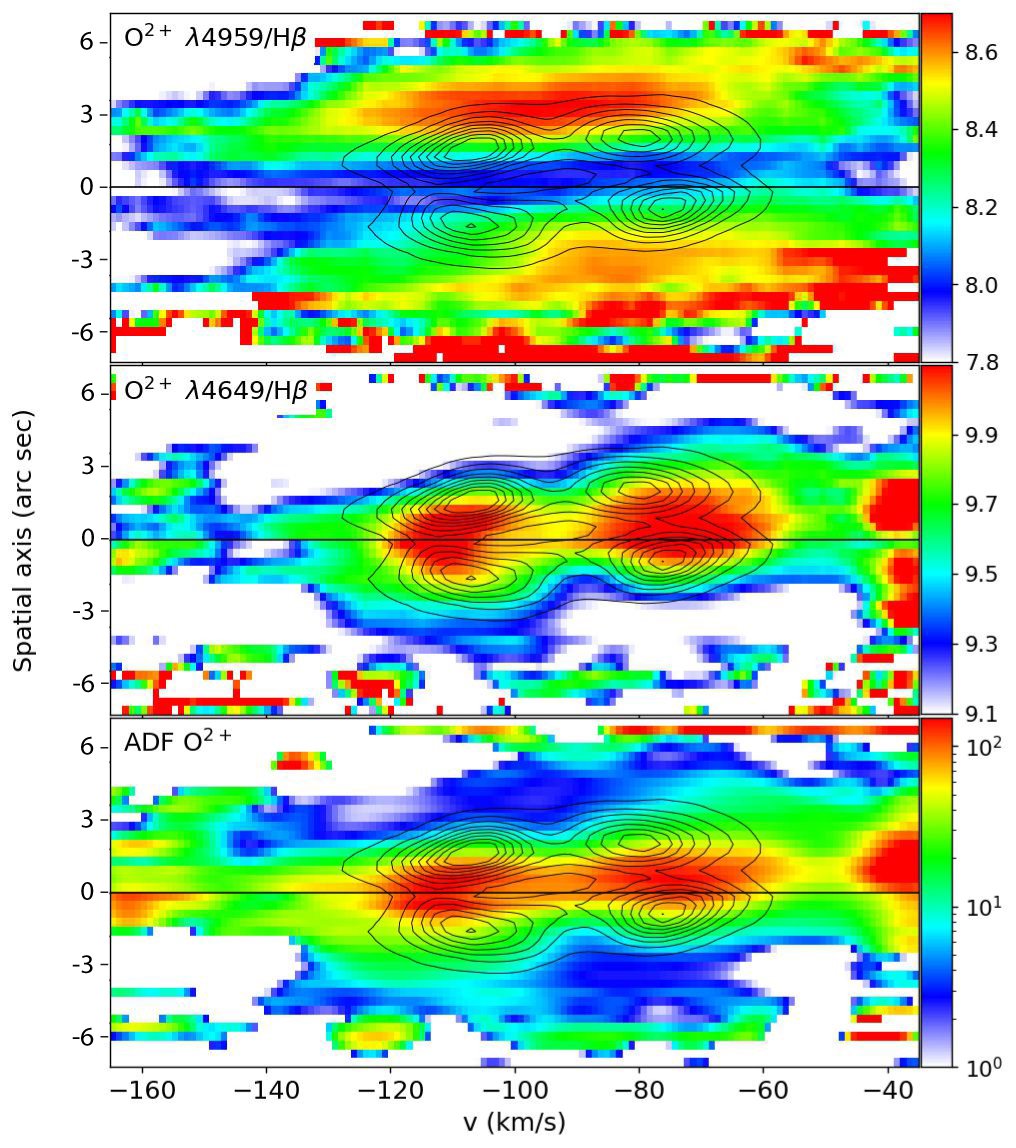}
    \caption{These PV diagrams present the $\mathrm O^{2+}/\mathrm H^+$ ionic abundance ratios for M 1-42 based upon the [\ion{O}{iii}] $\lambda$4959 (CEL, top) and \ion{O}{ii} $\lambda$4649 lines (RL, middle) in the top two panels.  The bottom panel displays the resulting PV diagram of the ADF for O$^{2+}$.  The PV diagrams of the abundances use the usual $12+\log(\mathrm O^{2+}/\mathrm H^+)$ scale while that for the ADF uses a logarithmic scale.  The contour lines correspond to the intensity of the \ion{O}{ii} $\lambda$4649 line.  The apparent increase at the very right side of the middle and bottom panels is due to emission from the \ion{C}{iii} $\lambda$4950 line and should be ignored.}
    \label{fig:oiiabundances-adf-m142}
\end{figure}

To compute the ionic abundances in M 1-42, we follow the same procedure described for Hf 2-2 (\S\ref{sec_Hf22_ADF}), though we broaden the \ion{O}{ii} $\lambda$4649 and [\ion{O}{iii}] $\lambda$4659 lines assuming temperatures of 1,000\,K and 7,500\,K, respectively. When computing the ionic abundances, for the normal nebular plasma, we adopt an electron density of 1,000\,cm$^{-3}$ and the electron temperature from the PV diagram of the [\ion{O}{iii}] temperature (Figure \ref{fig:tem[oiii-nii]-m142}).  For the additional plasma component, we adopt an electron temperature of 1,000\,K and an electron density of 5,000\,cm$^{-3}$. Using PyNeb's \texttt{getIonAbundance} function, we calculate the abundance ratios for each plasma component point by point across the PV diagrams. Figure \ref{fig:oiiabundances-adf-m142} presents the results based upon the [\ion{O}{iii}] $\lambda$4959 and \ion{O}{ii} $\lambda$4649 lines in the top and middle panels, respectively.  The morphologies and values of the abundance ratios based upon the two lines are very different.  

The asymmetry observed in the top panel of Figure \ref{fig:oiiabundances-adf-m142} is expected.  Figure 5 of \citet{garciarojas2022} indicates that $c(\mathrm H\beta)$ in the north-west part of M 1-42 is about 0.1 dex greater than in the south-east, which our spatially-integrated reddening cannot take into account.  The effect of such a difference on the [\ion{O}{iii}] $\lambda\lambda$4363,4959 lines would imply a temperature $\sim 500$\,K too low in the north-west and so an abundance that is $\sim 0.1$\,dex too large.  This systematic error accounts for most of the observed asymmetry.  The rest is presumably due to spatial differences in the ionization structure of O$^{2+}$ in M 1-42.

We compute the ADF map for M 1-42 by taking the ratio of the $\mathrm O^{2+}/\mathrm H^+$ abundance ratio maps from the ORL and CEL diagnostics. We present the result in Figure \ref{fig:oiiabundances-adf-m142}.  Given the different morphologies of the $\mathrm O^{2+}/\mathrm H^+$ abundance ratio maps, there is a very notable variation of the ADF across the nebula.  We find very high local values of the ADF along lines of sight near the central star, $\mathrm{ADF} > 100$, and values as low as $3-5$ along lines of sight towards the NW and SE edges of the object.  The lowest values of the ADF are for the most blue-shifted matter towards the NW and the most red-shifted matter towards the SE, which is the outermost matter in the approaching and receding lobes, respectively, if the orientation is as indicated in Figure \ref{fig:lobulos}. 

\subsection{Relative masses of N$^{2}$ and O$^{2}$ in the plasma components}

We employ the same methodology as in Hf 2-2, utilizing the decomposition of \ion{N}{ii} \lm5680 and \ion{O}{ii} \lm4649 (Figure \ref{fig:m142-nii-oii-decom}) to calculate the mass fraction of the additional plasma component relative to the normal plasma component in M 1-42. We adopt the physical conditions indicated in the previous section: the \te is set at [\ion{O}{iii}] PV map (Figure \ref{fig:tem[oiii-nii]-m142}, right), and the \nel is fixed at 1,000 cm$^{-3}$ for the normal nebular plasma; for the additional plasma component, the \te is set at 1,000 K and the \nel is fixed at 5,000 cm$^{-3}$.

\begin{figure*}
    \centering
    \includegraphics[width=0.89\linewidth]{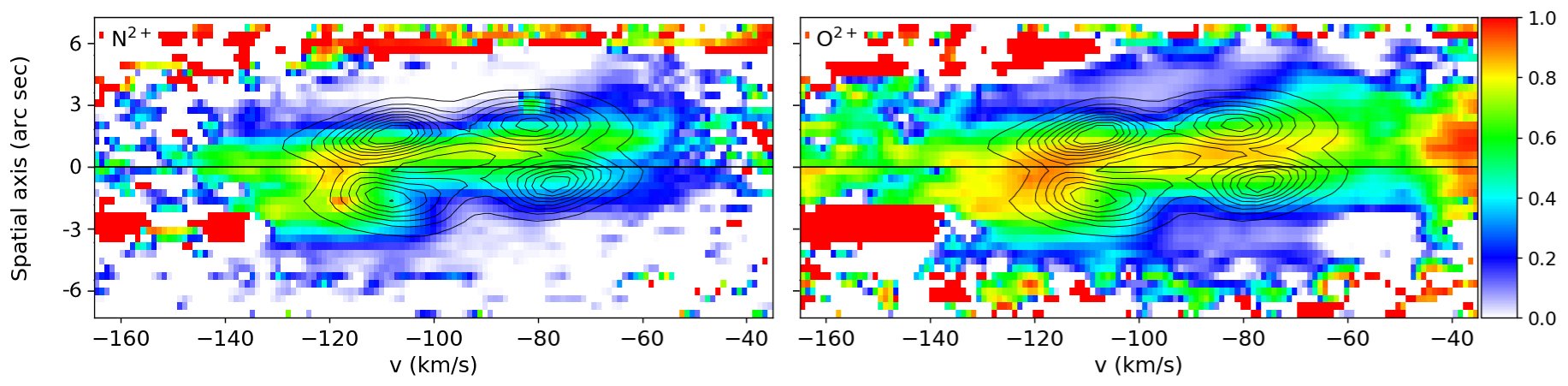}
    \caption{These PV diagrams of M 1-42 present the mass fraction of the N$^{2+}$ (left) and O$^{2+}$ ions (right) in the additional plasma component, assuming the physical conditions of (\te, \nel) = (1,000 K, 5,000 cm$^{-3}$) for the additional plasma component and (\te, \nel) = ([\ion{O}{iii}] \te PV diagram, 1,000 cm$^{-3}$) for the normal nebular plasma. The contour lines correspond to the intensity of the \ion{N}{ii} $\lambda$5680 (left) and \ion{O}{ii} $\lambda$4649 line (right).}
    \label{fig:relative-mass-m142}    
\end{figure*}

Figure \ref{fig:relative-mass-m142} presents the resulting PV diagrams of the mass fractions of N$^{2+}$ and O$^{2+}$ in the additional plasma component in M 1-42.  The morphologies of the two PV diagrams are very similar to the ADF map in Figure \ref{fig:oiiabundances-adf-m142}, but show finer details since the underlying PV diagrams are not broadened.  The mass fraction in the additional plasma component in both ions peaks along lines of sight near the central star and falls to zero in the disc along lines of sight farthest from the central star.  There is apparently a greater mean mass fraction for O$^{2+}$ than for N$^{2+}$.  However, for both ions, the additional plasma component contributes up to 80\% or more of the total mass of both ions.  The difference presumably implies that the two plasma components have different $\mathrm N^{2+}/\mathrm O^{2+}$ ratios or ionization structures.  

\begin{table*}
\caption{Summary of main results.}
\begin{tabular}{llll}
\hline
       &  & Indicator                           & Result                                                          \\ \hline
Hf 2-2 & $T_\mathrm{e}$   & [\ion{N}{ii}] $\lambda$$\lambda$5755/6583  & $\sim 8,000$\,K (outer shell), $\sim 10,000$\,K (inner shell)                   \\
       &           & [\ion{O}{iii}] $\lambda$$\lambda$4363/4959 & $\sim 8,000$\,K (outer shell), $\sim 10,000$\,K (inner shell), $16,000$\,K (near CS)                                                    \\
       &           & \ion{N}{ii} $\lambda$$\lambda$4041/5680    & $< 3,200$\,K \\
       &           &\ion{O}{ii} $\lambda\lambda$4089/4649      & $< 3,200$\,K \\
       & $n_\mathrm{e}$   & [\ion{S}{ii}] $\lambda$$\lambda$6731/6716  & $500$\,cm$^{-3}$ (outer shell), $\sim 2,000$\,cm$^{-3}$ (inner shell) \\
       &           &[\ion{O}{ii}] $\lambda\lambda$3729/3726   & $500-1,000$\,cm$^{-3}$ (outer shell), $\sim 2,000$\,cm$^{-3}$ (inner shell), $>3,000$\,cm$^{-3}$ (near CS) \\
       &           & \ion{N}{ii} $\lambda$$\lambda$5666/5680    & $> 3,000$\,cm$^{-3}$                                            \\
       &           &\ion{O}{ii} $\lambda$$\lambda$4662/4649    & $1,500-5,000$ cm$^{-3}$                                              \\
              &           &                                     &                   \\
M 1-42 & $T_\mathrm{e}$   & [\ion{N}{ii}] $\lambda$$\lambda$5755/6583  & $\sim 8,000$\,K (outer), $\sim 12,000$\,K (near CS) \\

       &           & [\ion{O}{iii}] $\lambda$$\lambda$4363/4959 & $\sim 7,000$\,K (outer), $13,000$\,K (near CS) \\
       &           & \ion{N}{ii} $\lambda$$\lambda$4041/5680    & $\sim 1,000$\,K \\
       &           &\ion{O}{ii} $\lambda\lambda$4089/4649      & $600-3,000$\,K \\
       & $n_\mathrm{e}$   & [\ion{S}{ii}] $\lambda$$\lambda$6731/6716  & $> 3,000 \,\mathrm{cm}^{-3}$ (inner peaks), $\sim 500\,\mathrm{cm}^{-3}$ (outer regions) \\
       &           &[\ion{O}{ii}] $\lambda\lambda$3729/3726   & $> 4,000\,\mathrm{cm}^{-3}$ (near CS), $\sim 500\,\mathrm{cm}^{-3}$ (outer regions) \\
       &           & \ion{N}{ii} $\lambda$$\lambda$5666/5680    & $> 3,000\,\mathrm{cm}^{-3}$                                          \\
       &           &\ion{O}{ii} $\lambda$$\lambda$4662/4649    & $> 4,000$\,cm$^{-3}$ and $T_e > 400\,\mathrm K$             \\ \hline
\end{tabular}
\label{summary}
\end{table*}

\section{Discussion}

This study has revealed both similarities and distinctive features in the kinematics and physical conditions of the planetary nebulae Hf 2-2 and M 1-42. The identification of two plasma components—normal and additional—in both objects underscores the complexity of their internal structures. However, each nebula exhibits unique characteristics that provide further insights into their evolution and ionization properties.

In both Hf 2-2 and M 1-42, the kinematics of the \ion{O}{i}, \ion{C}{ii}, \ion{N}{ii}, \ion{O}{ii} and \ion{Ne}{ii} lines differ from the kinematics of the other emission lines (Figures \ref{fig:wd-hf22} and \ref{fig:m142-kinematical}).  This result may be interpreted as indicating that these lines arise from an additional plasma component that emits in these permitted lines, but not (or very little) in the \ion{H}{i}, \ion{He}{i}, and forbidden lines.  In both Hf 2-2 and M 1-42, the electron density and temperature derived from the \ion{N}{ii} and \ion{O}{ii} lines differ systematically from those derived from forbidden line ratios (\S\S \ref{sec_hf22_physcond_permitted},\ref{sec_m142_physcond_permitted}), with the electron temperature being lower and the electron density higher.  Hence, the physical conditions derived from the permitted and forbidden lines also favour the interpretation of the presence of two plasma components in both objects.

The temperature gradients in the normal nebular plasma in Hf 2-2 and M 1-42 are much more dramatic than in NGC 7009 and NGC 6153 (Figures \ref{fig:tem_cel_hf22} and \ref{fig:tem[oiii-nii]-m142}).  Unfortunately, the [\ion{O}{iii}] temperature is the only temperature diagnostic available in the inner part of the O$^{2+}$ zone.  The volume of the nebula at the highest temperatures is presumably small and confined to the inner part of the O$^{2+}$ zone, because it does not appear in the [\ion{S}{iii}] temperature maps of \citet{garciarojas2022}.  The ionization energy of S$^{2+}$ is similar to that for O$^+$, so the [\ion{S}{iii}] temperature should not be very sensitive to the inner part of the O$^{2+}$ zone.  If real, these strong temperature gradients imply unusually strong heating in this volume in Hf 2-2 and M 1-42.  

In Hf 2-2, the additional plasma component has spatial and velocity distributions that imply that it is internal to the normal nebular plasma (Figure \ref{fig:oiiabundances-adf-hf22}), in agreement with the spatial distribution found by \citet{garciarojas2022}.  In M 1-42, the additional plasma component is similarly ``confined" in the sense that it appears to be contained within the emission that defines the approaching and receding lobes (Figures \ref{fig:lobulos} and \ref{fig:oiiabundances-adf-m142}).  The same configuration, of the additional plasma component enclosed by the normal nebular plasma also holds in NGC 6153 and NGC 7009 \citep{richer2013ApJ, richer2022ngc}.  Thus, in these four objects with large ADFs, there is a notable difference in the spatial distribution of the two plasma components, with the additional plasma component found closer to the central star.  This also agrees with the well-known result that the emission from the permitted lines of \ion{C}{ii}, \ion{N}{ii}, and \ion{O}{ii} is more compact spatially than the forbidden emission from \ion{C}{iii}] or [\ion{O}{iii}] \citep[e.g.,][]{barker1982, garnettdinerstein2001, tsamisetal2008, garciarojasetal2016}.  

The larger velocities measured for the additional plasma component in M 1-42 is an interesting result (Figure \ref{fig:m142-kinematical}).  It recalls the models of \citet{FrankMelemaIV1994} for axially symmetric planetary nebulae whose velocity varies as a function of the polar angle.  In their model, the additional plasma component, concentrated at smaller polar angles, has a larger velocity than the normal nebular plasma found at lower latitudes towards the equator, as observed in M 1-42.  

The ADF varies enormously within Hf 2-2 and M 1-42.  The minimum value in M 1-42 is near unity, but is only $2-4$ in Hf 2-2.  In both objects, the ADF achieves maximum values exceeding a factor of 100 along lines of sight close to the central stars and, in Hf 2-2, at velocities close to the systemic velocity (Figures \ref{fig:oiiabundances-adf-hf22} and \ref{fig:oiiabundances-adf-m142}).  This agrees with the spatial variation found by \citet{garciarojas2022}.  The maximum values we find are greater than those found by \citet{garciarojas2022}, but that is expected since our results resolve the structure along the line of sight whereas their maps do not.   

It is clear from the middle panels of Figures \ref{fig:oiiabundances-adf-hf22} and \ref{fig:oiiabundances-adf-m142} that the $\mathrm O^{2+}/\mathrm H^+$ ratio does not follow the emission in the \ion{O}{ii} $\lambda$4649 line in either Hf 2-2 or M 1-42. This is clear evidence that the O$^{2+}$ ions that emit the majority of the \ion{O}{ii} $\lambda$4649 emission in these objects do not follow the distribution of the H$^+$ ions.  Hence, in both Hf 2-2 and M 1-42, the additional plasma component must be deficient in hydrogen since this plasma component emits the majority of the \ion{O}{ii} $\lambda$4649 emission.

Hf 2-2, like NGC 6153 and NGC 7009, has the kinematics of a closed shell in both permitted and forbidden lines (Figure \ref{fig:ion}).  In NGC 6153 and NGC 7009, the morphology of the PV diagrams of the ADF for O$^{2+}$ is also that of a closed shell \citep[][their Figures 3 and 39, respectively]{Richer2019ApJ, richer2022ngc}.  However, in Hf 2-2, the morphology of the PV diagram of the ADF for O$^{2+}$ is not shell-like, since the ADF continues to increase inside the inner shell (Figure \ref{fig:oiiabundances-adf-hf22}).  Granted, the mass of the matter responsible for the emission that produces the very high ADF at the central position and velocity must be small given that it emits little in either permitted or forbidden lines (Figure \ref{fig:oiiabundances-adf-hf22}).  

A similar situation occurs in M 1-42, where the ADF also achieves values in excess of 100 along the line of sight towards the central star (Figure \ref{fig:oiiabundances-adf-m142}).  The positions and velocities of the two maxima of the ADF correspond to the interior of the lobes in M 1-42 (compare with Figure \ref{fig:lobulos}).  As in Hf 2-2, these maxima do not coincide with the maximum emission in the permitted lines.  This suggests that the difference noted above between Hf 2-2 and NGC 6153/NGC 7009 may reflect a different spatial relationship between the two plasma components in objects with the largest ADFs. 

The mass fractions of O$^{2+}$ ions in the additional plasma component that we find for M 1-42 and Hf 2-2 exceed those we have previously found in NGC 7009 and NGC 6153 \citep{Richer2019ApJ, richer2022ngc}.  This is perhaps not surprising given the larger ADFs in M 1-42 and Hf 2-2.  The maximum values of the mass fraction occur along lines of sight close to the central stars and, in Hf 2-2, at velocities close to the systemic velocity (Figures \ref{fig:relative-mass-hf22} and \ref{fig:relative-mass-m142}).  Along these lines of sight in both objects, the mass fraction of the additional plasma component approaches or exceeds 90\% of the total mass of N$^{2+}$ and O$^{2+}$ ions.  In both Hf 2-2 and M 1-42, the minimum mass fraction of the  N$^{2+}$ and O$^{2+}$ ions in the additional plasma component is approximately 50\% over almost all of the area where there is significant emission from [\ion{O}{iii}] $\lambda$4959 (Figures \ref{fig:relative-mass-hf22} and \ref{fig:relative-mass-m142}).  Hence, for the volume seen by the spectrograph slit in both objects, the additional plasma component contains at least 50\% of the total mass in both ions.  

The mass fractions we find here for the additional plasma component are higher than those found by \citet{garciarojas2022} in both Hf 2-2 and M 1-42.  In part, this is due to differences in the adopted physical conditions, but our results are also higher because of the larger fractional contribution from the central regions in our more limited spatial coverage (the spectrograph's slit).  

Evidently, when there is an additional plasma component, any study of the chemical composition of the object in question must account for it.  In the case of objects like Hf 2-2 or M 1-42, the additional plasma component contributes very significantly, at least for ions such as N$^{2+}$ and O$^{2+}$ (Figures \ref{fig:relative-mass-hf22} and \ref{fig:relative-mass-m142}).  For objects with lower ADFs, the contribution from the additional plasma component is still important \citep{Richer2019ApJ, richer2022ngc}.  However, it is not simply a matter of scaling the abundance from the normal nebular plasma by the ratio of mass fractions.  That procedure will yield only a lower limit to the relative abundance since such scaling ignores the hydrogen content of the additional plasma component.  (In the end, we seek $\mathrm O^{2+}/\mathrm H^+$.)  Correctly accounting for this contribution is complicated, because it is very difficult to account for the hydrogen found in the additional plasma component \citep{GomezLlanosMorisset2020, richer2022ngc, garciarojas2022, gomezllanosetal24}.  The foregoing considers only one ionization stage.  It is now evident that there are abundance discrepancies in O$^+$ \citep[e.g.,][]{garciarojasetal2016,garciarojas2022}, which may well be underestimated given the contribution that recombination can make to the [\ion{O}{ii}] lines, as occurs here for Hf 2-2 and M 1-42.  When the physical conditions and chemical compositions differ in the two plasma components, it is likely that their ionization structure will also differ \citep[e.g.,][]{GomezLlanosMorisset2020}.  Given the distinct physical conditions, ionization structures, and chemical compositions in multiple plasma components, obtaining a complete account of the chemical inventory of a nebula is a complex task.  

It is interesting to contrast the ADFs we obtain for Hf 2-2 and M 1-42 with that for NGC 6153 \citep[\S\ref{sec_Hf22_ADF}, \S\ref{sec_m142ADF},][]{richer2022ngc}.  We express the line intensity, $I(\lambda)$, due to ion $X$ as
\begin{equation}\label{eq_line_intensity}
    I(\lambda)=N_e N(X) \epsilon(\lambda, T_e) V
\end{equation}
\noindent where $N_e$ and $N(X)$ are the densities of electrons and ions, $T_e$ is the electron temperature, $\epsilon(\lambda, T_e)$ is the line emissivity, and V is the volume that emits the line observed by the spectrograph slit.  We calculate the ADF following the definition of \citet{tsamis2004}, restricted to the O$^{2+}$ ion

\begin{equation}\label{eq_ADF_single_comp}
    ADF = \frac{n(\mathrm O^{2+})_{ORL}/n(\mathrm H^+)}{n(\mathrm O^{2+})_{CEL}/n(\mathrm H^+)}
    = \frac{\frac{I(4649)}{N_e \epsilon(4649, T_e)}/\frac{I(H\beta)}{N_e \epsilon(H\beta, T_e)}}{\frac{I(4959)}{N_e \epsilon(4959, T_e)}/\frac{I(H\beta)}{N_e \epsilon(H\beta, T_e)}}\, ,
\end{equation}
  
\noindent noting that the number of ions is $N(X) V$.  If we assume a single plasma component, the terms involving H$\beta$ cancel.  Thus, the ADF is a function of the line intensities and the physical conditions in the plasma (Eq. \ref{eq_ADF_single_comp}).  Considering the physical conditions for either the normal nebular plasma or the additional plasma component in Table \ref{summary}, we see that they are not dramatically different from those for NGC 6153 \citep[][Table 9]{richer2022ngc}.  Yet, M 1-42 and Hf 2-2 have ADFs that are much higher, by factors of 2.5 and 7, respectively.  This suggests that the physical conditions are not primarily responsible for the ADFs, but rather the mass of O$^{2+}$ present.  

This conclusion is even more evident if we assume two plasma components, since then 
\begin{equation}
    ADF = \frac{n_n(\mathrm O^{2+})_{ORL}/n_n(\mathrm H^+)+n_a(\mathrm O^{2+})_{ORL}/n_a(\mathrm H^+)}{n_n(\mathrm O^{2+})_{CEL}/n_n(\mathrm H^+)+n_a(\mathrm O^{2+})_{CEL}/n_a(\mathrm H^+)}\, ,
\end{equation}
\noindent where the subscripts $n$ and $a$ indicate the normal nebular plasma and the additional plasma component, respectively.  Here, the two terms involving the normal nebular plasma must yield the same $\mathrm O^{2+}/\mathrm H^+$ ratio if ionization and thermal equilibria hold, since both terms involve exactly the same O$^{2+}$ ions.  Given the low temperature of the additional plasma component, any forbidden emission from the additional plasma component will be exceedingly difficult to separate from that due to the normal nebular plasma, and so is likely to be approximated as zero.  However, the second term in the numerator, the permitted emission from the additional plasma component will not be zero, as it emits the majority of the permitted emission.  The difficulty in computing this term is in estimating the H$\beta$ emission associated with the additional plasma component \citep{GomezLlanosMorisset2020, garciarojas2022, richer2022ngc, gomezllanosetal24}.  In any case, it is clear that this term sets the value of the ADF and it depends upon the mass of O$^{2+}$ present in the additional plasma component.  

Thus, the larger ADFs in M 1-42 and Hf 2-2 compared to NGC 6153 would appear to arise due to intrinsically larger masses of O$^{2+}$ ions in these objects, not the physical conditions.  If this situation holds generally, the ADF arises as a result of a variable mass of O$^{2+}$ ions that emits in permitted lines, but not in forbidden lines.  

The foregoing assumes that any complications related to the physical conditions are taken into account (e.g., temperature fluctuations).  Also, were it feasible to reliably measure the forbidden emission from the additional plasma component, $ADF=1$ should result.  The foregoing should also illustrate why the ADF  "problem" is a problem of the analysis of the data.  

\section{Conclusions}

This study highlights the kinematic and physical properties of Hf 2-2 and M 1-42, two planetary nebulae with high ADFs, of 70 and 20, respectively \citep{liuetal2001, liu2006}. As we have found previously in NGC 7009 and NGC 6153, both Hf 2-2 and M 1-42 apparently contain two plasma components with distinct kinematics and physical conditions.  In both Hf 2-2 and M 1-42, the normal nebular plasma has gradients in both the electron density and temperature.  On the other hand, the physical conditions in the additional plasma component are more uniform, with much lower electron temperatures and significantly higher electron densities.  

In Hf 2-2, the majority of the emission lines present the typical trend of increasing line splitting as the ionization potential of the parent ion decreases \citep{wilson1950ApJ}. These lines define the normal nebular plasma.  On the other hand, the lines of \ion{O}{i}, \ion{C}{ii}, \ion{N}{ii}, \ion{O}{ii}, and \ion{Ne}{ii} vary very little as a function of the ionization potential, which defines the additional plasma component.  The physical conditions in these two plasma components are likewise distinct.  In the normal nebular plasma, the electron density is lowest, of order $500-1,000$\,cm$^{-3}$, in the outer shell, but rises to $\sim 2,000$\,cm$^{-3}$ closer to the central star.  The electron temperature presents a very strong temperature gradient, from 8,000\,K in the outer shell, 10,000\,K in the inner shell, and rising to 16,000\,K in a small volume near the central star.  In contrast, the additional plasma component has a uniform electron density and temperature with values of $3,000-5,000$\,cm$^{-3}$ and $< 3,200$\,K, respectively. 

In M 1-42, the analysis of its kinematics is more complex due to its bipolar morphology and our viewing angle close to the polar axis.  Most of the matter appears to be in a toroid-like structure, so, except in the low ionization lines, the emission arises mostly from the front and back sides of the toroid.  Hence, we study the kinematics using only the higher ionization lines and determine the motion away from the systematic velocity.  We again find two kinematic components, but with the roles reversed compared to Hf 2-2.  In M 1-42, most of the lines have very similar motions with respect to the systemic velocity, which is how we define the normal nebular plasma.  In contrast, the lines of \ion{O}{i}, \ion{C}{ii}, \ion{N}{ii}, \ion{O}{ii}, and \ion{Ne}{ii} present a systematic variation as a function of the ionization potential and define the kinematics of the additional plasma component.  These lines show a smaller spatial extent than the lines from the normal nebular plasma.  However, deep PV diagrams of the [\ion{O}{iii}] $\lambda$4959 and [\ion{N}{ii}] $\lambda\lambda$6548,6583 lines indicates that the additional plasma component is contained within the lobes defined by these lines.  The electron density and temperature in the normal nebular plasma vary substantially, from $\sim 500\,\mathrm{cm}^{-3}$ and $\sim 7,000$\,K in the outer parts of the nebula to over $3,000\,\mathrm{cm}^{-3}$ and $13,000$\,K closer to the central star.  In the additional plasma component, the  electron temperature and density appear to vary less, with values of $> 4,000$\,cm$^{-3}$ and $\sim 1,000$\,K.  

We decompose the permitted emission in the \ion{O}{ii} $4649$ and \ion{N}{ii} $5680$ lines and compute PV diagrams of the ADF and the mass fraction of O$^{2+}$ and N$^{2+}$ ions in the additional plasma component.  Both vary strongly, with the ADF achieving values in excess of 100 while the mass fraction approaches or exceeds 90\% of the total mass of these ions in the central parts (projected) in both Hf 2-2 and M 1-42.  In objects such as these, it is crucial to include the contribution of the additional plasma component when computing the total chemical composition.  However, doing so is difficult, as others have also emphasized.  

\section{Acknowledgements}

Based on observations collected at the European Southern Observatory under ESO programme ID 69.D-0174(A).  We gratefully acknowledge funding from DGAPA-UNAM grants IG101233 and IN111124.  LCCC gratefully acknowledges support from a Secretar\'ia de Ciencia, Humanidades, Tecnolog\'ia e Innovaci\'on scholarship.  

\section*{Data Availability}

The individual PV images will be  available at the VizieR database.



\bibliographystyle{mnras}
\bibliography{mnras-articulo/ref} 




\appendix 

\section{}

\begin{figure*}
    \
    \includegraphics[width=0.89\linewidth]{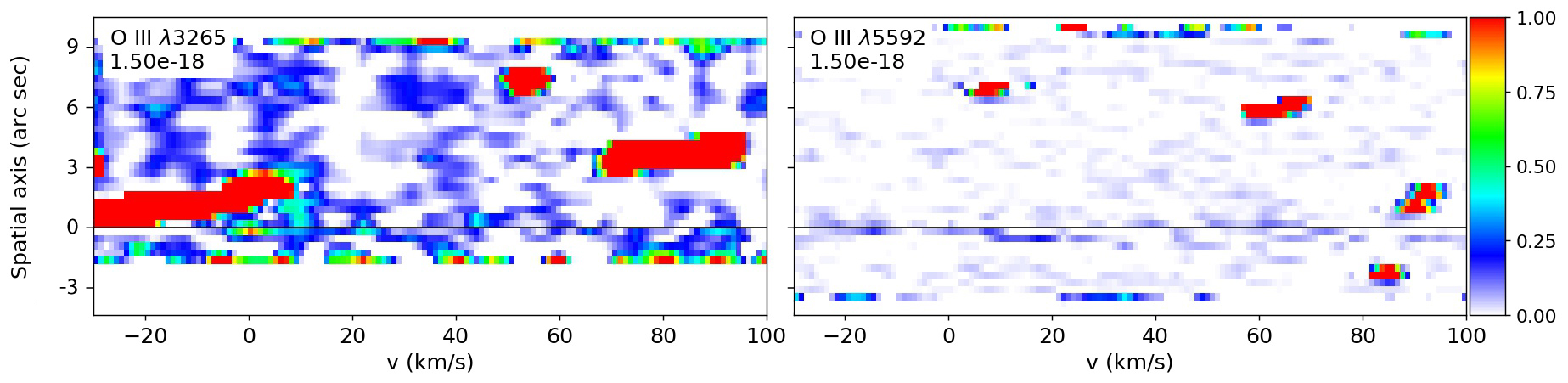}
    \caption{We present the PV diagrams for Hf 2-2 for the permitted lines of \ion{O}{iii} $\lambda$3265 and $\lambda$5592, due to recombination and charge exchange, respectively. The lack of emission indicates that there is no significant contribution to the [\ion{O}{iii}] lines due to these mechanisms.}
    \label{fig:reclines}
\end{figure*}

\begin{figure*}
    \centering
    \includegraphics[width=0.89\linewidth]{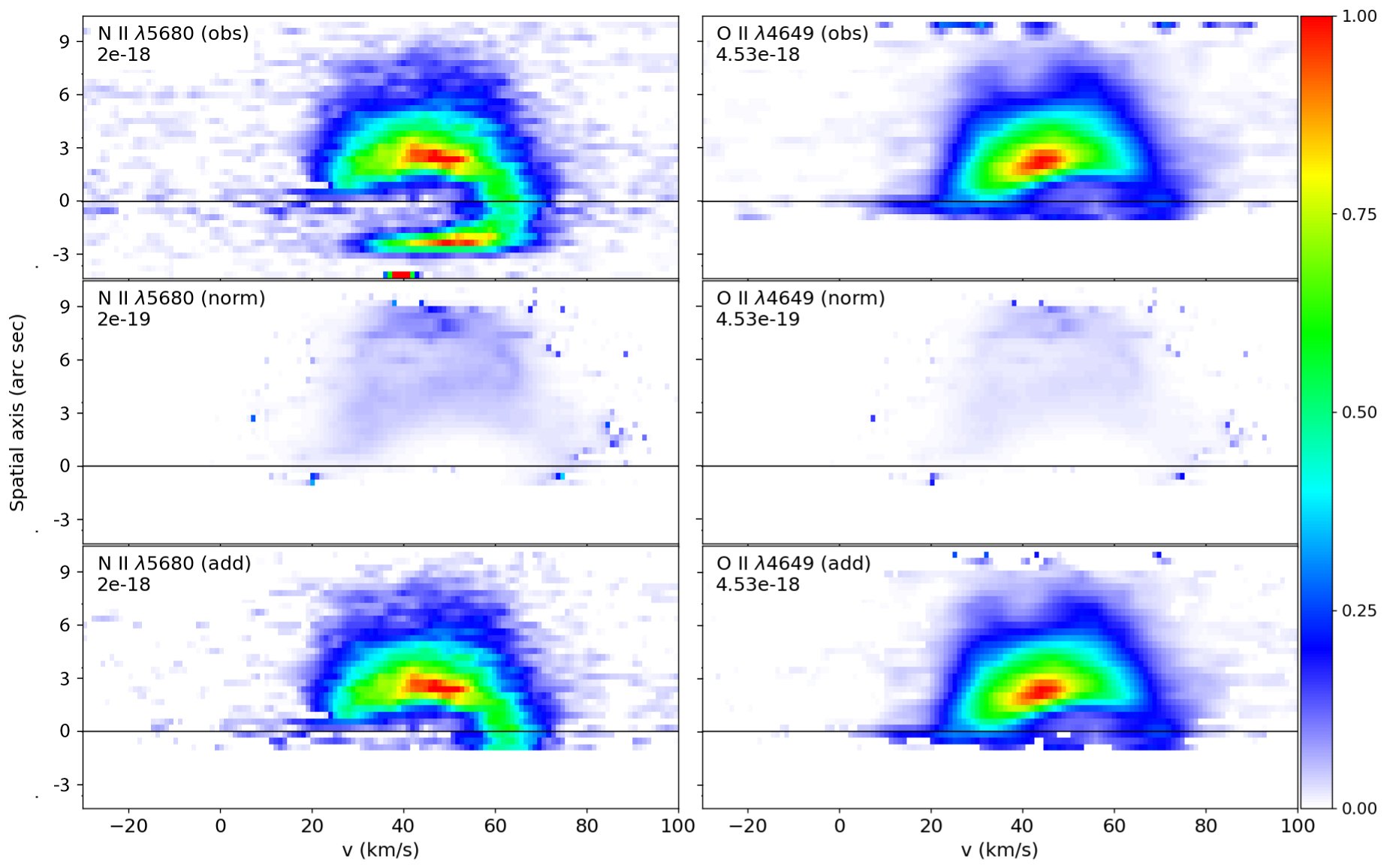}
    \caption{These PV diagrams show the decomposition of \ion{N}{ii} $\lambda$5680 (left) and \ion{O}{ii} $\lambda$4649 lines (right) in Hf 2-2. For both species: the top panel displays the observed line, the middle panel shows the contribution from the normal nebular plasma, and the bottom panel presents the additional plasma component, which is the difference between the two previous panels.  Note that the scale factor in the second row differs from that in the other two in order for the emission predicted for the normal nebular plasma to be visible.  Following \citet{richer2022ngc}, we model the emission from the normal nebular plasma as a scaled version of the PV diagram of the [\ion{O}{iii}] $\lambda$4959 line.  We assume an abundance ratio $\mathrm N^{2+}/\mathrm O^{2+} = 0.46$ \citep{liu2006}.}
    \label{fig:hf22-nii-oii-decom}
\end{figure*}

\begin{figure*}
    \centering
    \includegraphics[width=0.89\linewidth]{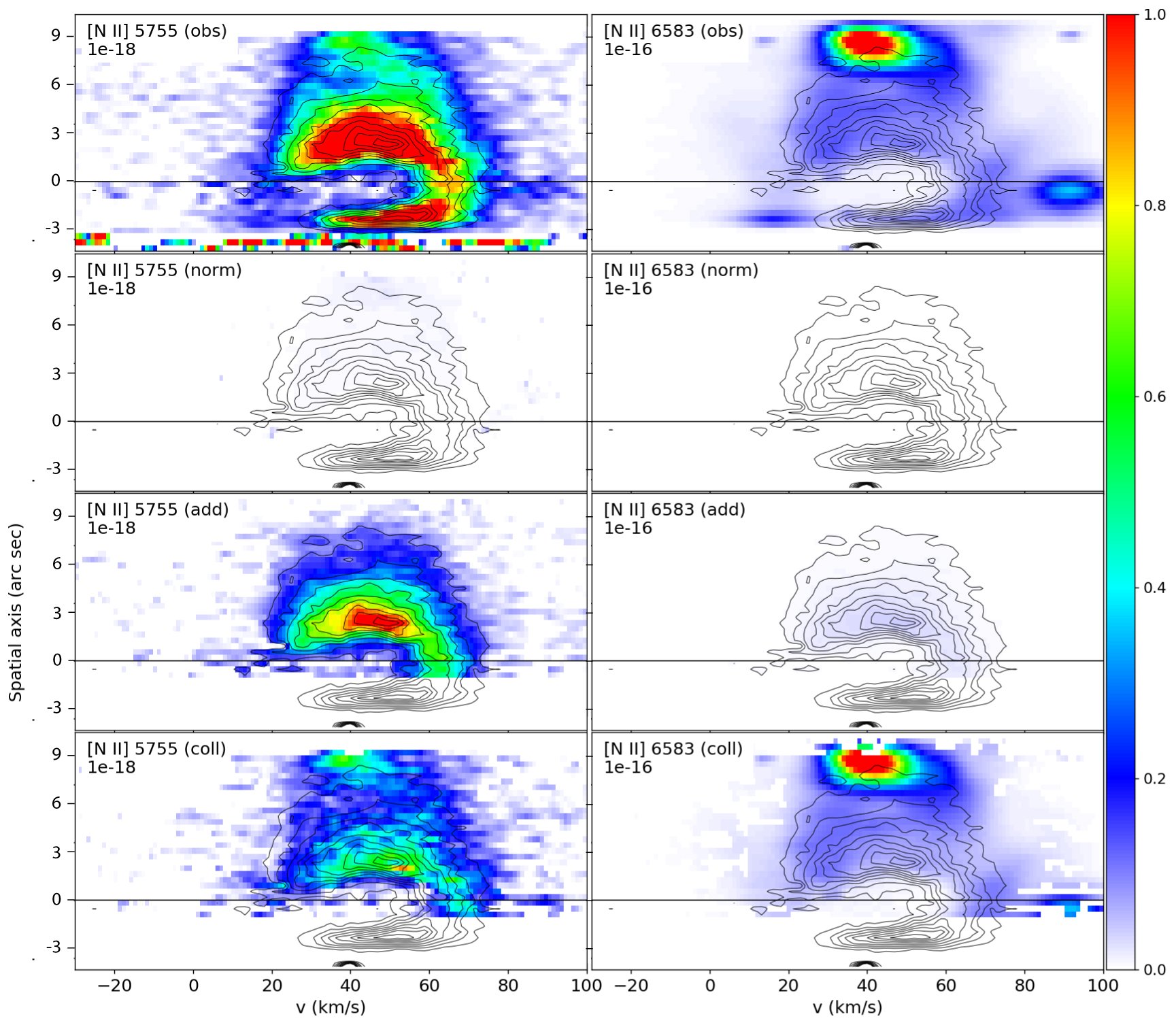}
    \caption{These PV diagrams illustrate the subtraction of the recombination contribution to the [\ion{N}{ii}] $\lambda\lambda$5755,6583 lines in Hf 2-2. In the top row, we present the observed PV diagrams of these lines. The second and third rows show the modelled permitted emission in these lines due to recombination in the normal nebular plasma and in the additional plasma component, respectively. The bottom row presents the residual emission after subtracting the emission in the two previous panels from the observed PV diagram, i.e., what should be the collisional component of the forbidden lines. In all cases, contours of the intensity of the \ion{N}{ii} $\lambda$5680 line intensity is overplotted for reference.  Following \citet{richer2022ngc}, we model the recombination contribution from the normal nebular plasma using a scaled PV diagram of [\ion{O}{iii}] $\lambda$4959 and we adopt the model due to recombination in the additional plasma component using the decomposition in Figure \ref{fig:hf22-nii-oii-decom}.}
    \label{fig:hf22-decon-nii}
\end{figure*}

\begin{figure*}
    \centering
    \includegraphics[width=0.89\linewidth]{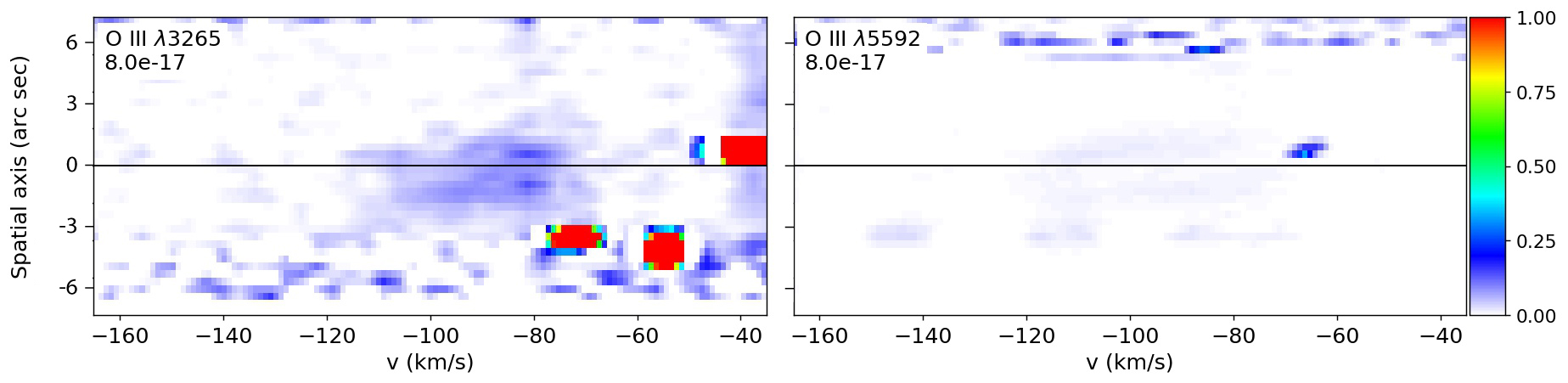}
    \caption{We present the PV diagrams for M1-42 of the emission in the permitted lines of \ion{O}{iii} $\lambda$3265 and $\lambda$5592 from recombination and charge exchange, respectively. The flux scale is the same as used for the [\ion{O}{iii}] $\lambda$4363 line (Figure \ref{fig:tem[oiii-nii]-m142}).  Given the faintness of emission at these wavelengths, we conclude that recombination and charge exchange contribute minimally to the [\ion{O}{iii}] $\lambda$4363 emission.}
    \label{fig:reclines-m142}
\end{figure*}

\begin{figure*}
    \centering
    \includegraphics[width=0.89\linewidth]{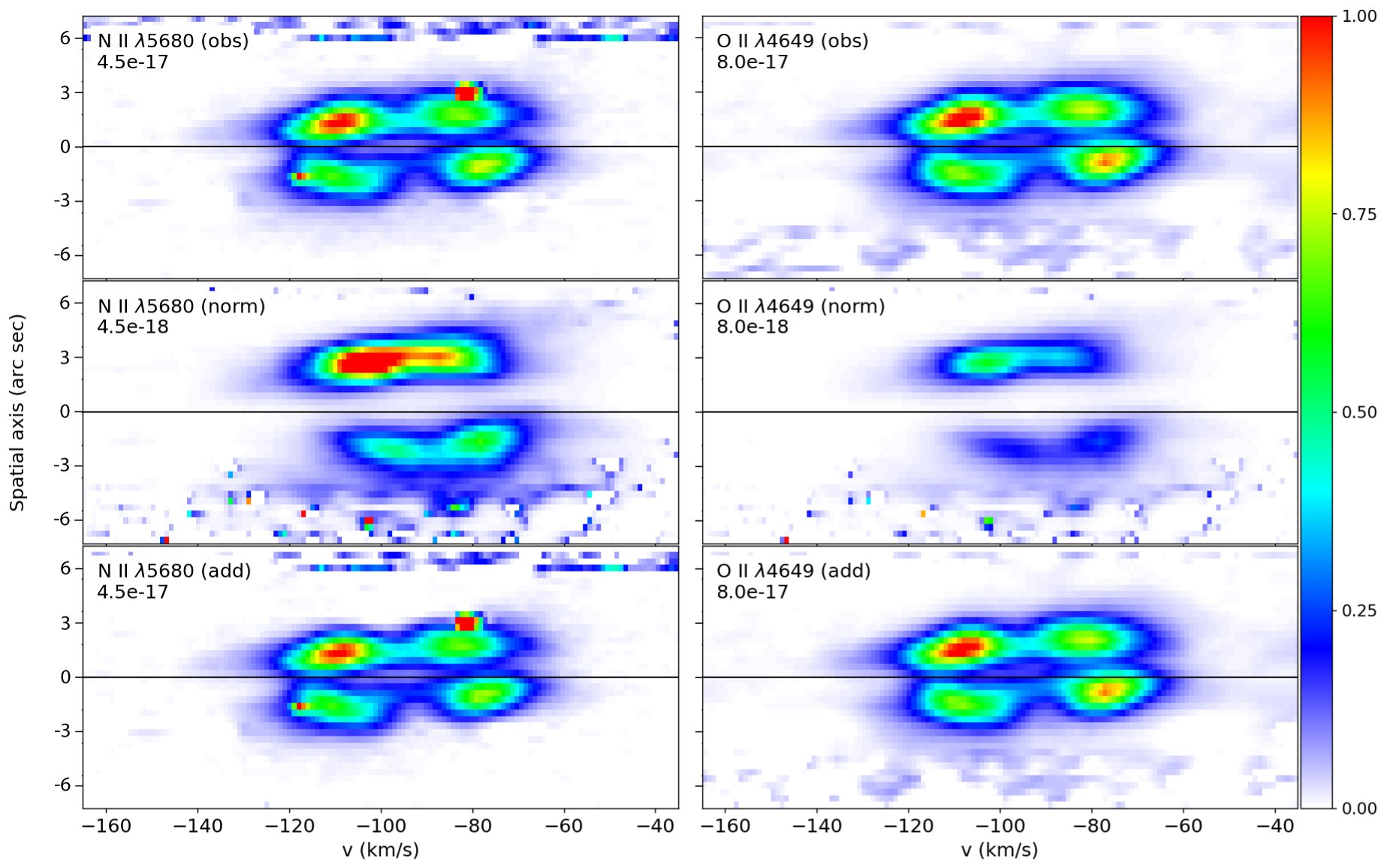}
    \caption{These PV diagrams show the decomposition of the \ion{N}{ii} $\lambda$5680 (left) and \ion{O}{ii} $\lambda$4649 (right) lines in M 1-42. For both ions, the top panel displays the observed emission, the middle panel shows the predicted emission from the normal nebular plasma, and the bottom panel presents the additional plasma component.  See Figure \ref{fig:hf22-nii-oii-decom} for further details.  We assume an abundance ratio $\mathrm N^{2+}/\mathrm O^{2+} = 1.13$ \citep{liuetal2001}.}
    \label{fig:m142-nii-oii-decom}
\end{figure*}

\begin{figure*}
    \
    \includegraphics[width=0.89\linewidth]{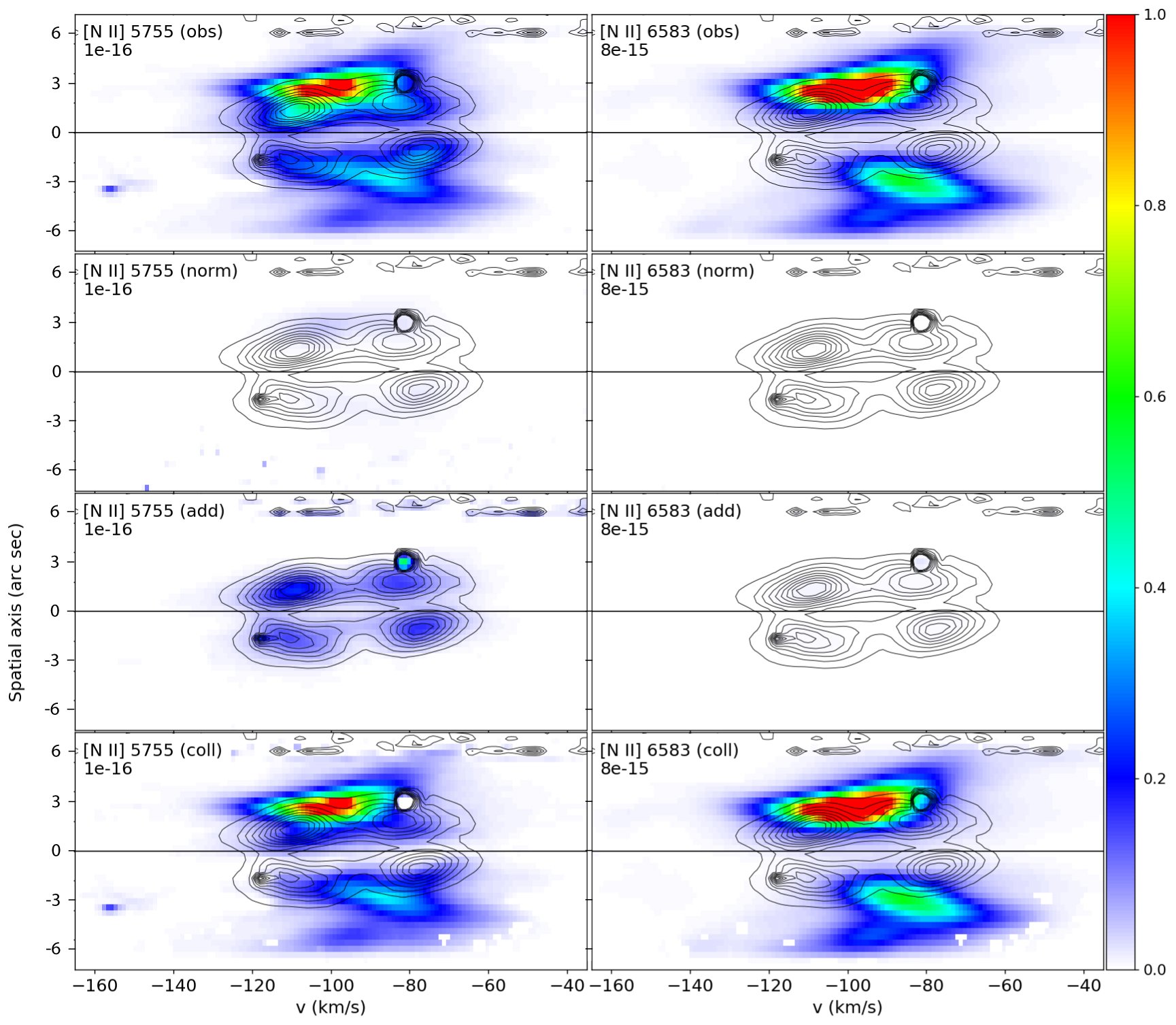}
    \caption{These PV diagrams illustrate the subtraction of the recombination contribution to the [\ion{N}{ii}] $\lambda\lambda$5755,6583  lines in M 1-42. In the top row, we present the observed PV diagrams of these lines. The second and third rows show the modelled permitted emission in these lines due to recombination in the normal nebular plasma and in the additional plasma component, respectively. The bottom row presents the residual emission, i.e., the expected collisional component of the forbidden lines, obtained after subtracting the modelled recombination contribution from the observed PV diagrams. In all cases, the contours of the \ion{N}{ii} $\lambda$5680 line intensity is overplotted for reference \citep[details:][]{richer2022ngc}.  See Figures \ref{fig:hf22-decon-nii} and \ref{fig:m142-nii-oii-decom} for further details.}
    \label{fig:m142-decon}
\end{figure*}


\bsp	
\label{lastpage}
\end{document}